\documentclass[11pt,a4paper]{article}
\pdfoutput=1
\usepackage{jheppub}

\usepackage{graphicx}
\usepackage{verbatim}
\newcommand{\ba}{\begin{eqnarray}}
\newcommand{\ea}{\end{eqnarray}}
\newcommand{\no}{\nonumber}
\newcommand{\be}{\begin{equation}}
\newcommand{\ee}{\end{equation}}
\newcommand{\bea}{\begin{eqnarray}}
\newcommand{\eea}{\end{eqnarray}}

\title{
WIMPs and Un-Naturalness
}
\date{\today
}
\author{
Luca Vecchi}
\affiliation{Maryland Center for Fundamental Physics,\\ Department of Physics, University of Maryland\\
College Park, MD 20742, USA}
\emailAdd{vecchi@umd.edu}
\abstract{The WIMP ``miracle" suggests a new physics threshold ranging from the weak scale up to several tens of TeVs. Obtaining the correct dark matter density in many theories aiming to solve the hierarchy problem may thus require some amount of tuning of the weak scale, hinting at a possible connection between WIMP dark matter and unnaturalness. We point out that dark matter direct detection is a very efficient probe of these unnatural models, and that existing data already provide important clues to the nature of the associated WIMPs. We present a model-independent, relativistic analysis of the signatures of a gauge-singlet dark matter candidate of arbitrary spin, and discuss the current experimental bounds from LUX and XENON100. For complex WIMPs, dark matter direct detection is complementary to electroweak precision tests, and can even compete with flavor constraints if the dark matter has spin. Particularly relevant for future searches are couplings to the Higgs mass operator, which are expected to be large if the electroweak scale is finely tuned. Care is devoted to the RG evolution of the effective Lagrangian. We find that the CP-even scalar coupling to charm quarks is enhanced by about 20\% compared to the one-loop estimate. 

When pushed in the unnatural regime, warped extra dimensions --- with or without custodial symmetry --- become attractive theories for flavor, the Higgs mass, and dark matter. The WIMP argument basically sets an upper bound on unnaturalness, whereas direct detection experiments select scalar or real particles as the most compelling dark matter candidates.
}
\begin{document}
\maketitle
%

\section{WIMPs and Un-Naturalness}

So far, the LHC has found no evidence of physics beyond the standard model (SM). One possible reason might be that the physics responsible for solving the hierarchy problem is somewhat different from what we were expecting. The alternative option is that the weak scale is tuned.

The most serious tuning in particle physics is associated with the cosmological constant, and to date, arguably the simplest solution to this puzzle is offered by ``environmental considerations"~\cite{Weinberg:1987dv}. Perhaps the big hierarchy is due to similar reasons. 

Yet, as opposed to the cosmological constant, there exist very compelling theories for the smallness of the weak scale which, as far as we know, would be able to consistently describe Nature up to the Planck scale. If not for null collider searches, there would be no apparent reason for them to hide in the multi-TeV range.

An optimistic viewpoint is that the new physics threshold is truly around the corner, say closer to $10$ TeV than the weak scale, and that this little hierarchy is anthropically motivated. Dark matter (DM) may play a central role in this picture. 

Realistic SUSY and strongly coupled theories for the weak scale generically have dark matter candidates~\cite{Vecchi:2013xra}, with the observed DM relic density typically obtained for masses $\sim g_X^2(3-5)$ TeV $\leq O(100)$ TeV.~\footnote{An upper limit may be obtained by rescaling low energy proton-antiproton scattering. The result is consistent with what one expects from a dynamics that saturates the non-perturbative bound $g_X\sim4\pi$.} The DM scale is above the TeV for couplings $g_X\sim1$ or larger, and this often implies that also the states immediately involved in the cancellation of the quadratic divergence in the Higgs mass are somewhat too heavy. In a model-building sense, these would-be natural theories must therefore ``choose" between two regimes: one with a new physics threshold above the TeV, right where the WIMP argument wants, and the other around a few hundred GeV, as suggested by naturalness. In the first case we would say that the theory is technically unnatural. However, the second class of models often has a hidden tuning in ``model space", as referred to in~\cite{ArkaniHamed:2012gw}, so it is difficult to quantify which alternative is more {\emph{likely}}. 

In other words, in the multiverse both the DM abundance and the weak scale would scan in model space, and the new physics scale would thus be subject to two competing ``pressures". One favors a large $\Omega_{\rm DM}/\Omega_{\rm b}\gtrsim5$ (see for instance~\cite{Hellerman:2005yi}), which in these models means a large new physics scale, and the latter pushes the theory towards naturalness~\cite{Agrawal:1997gf}. If the pull from the former happens to be stronger, then we should not be too surprised to find ourselves in a world with a finely-tuned weak scale.~\footnote{A similar logic has been discussed in~\cite{Bousso:2013rda} in the context of supersymmetric extensions. Our point of view is that this argument is more general, and perhaps even a bit sharper in warped extra dimensions, where the number of independent parameters that may scan is virtually smaller. In particular, the mass scale controlling naturalness and DM is the same.}

The WIMP ``miracle" now appears as the result of a fine-balancing in model-space, and the very existence of stable particles is at the origin of the unnaturalness of the weak scale. The connection between DM and unnaturalness is especially neat in models that do not admit a hierarchical structure in the spectrum of the new resonances, as it is in generic strongly coupled scenarios, in which case the WIMP argument truly sets an ``upper bound on unnaturalness". 

Within this picture, WIMP searches should be viewed as an independent probe of would-be natural theories for the weak scale, complementary to electroweak and flavor constraints.

Unfortunately, the WIMP ``miracle" does not say anything specific about the actual nature of DM beyond the magnitude of its annihilation cross section. WIMPs could be self-conjugate (real) as well as complex fields, could have a variety of gauge quantum numbers and spins, and masses spanning several orders of magnitudes. It is therefore of primary importance to study the phenomenology of WIMPs without making reference to any particular model.

Direct detection experiments offer a unique opportunity in this respect. Because the momentum transfer characteristic of elastic DM-nuclei scattering is comfortably below the DM mass, it is possible to use standard effective field theory techniques to perform an operator analysis. In contrast, indirect detection from DM annihilation and decay, as well as collider searches, probe energies comparable to the new physics scale, and are intrinsically sensitive to the details (i.e. new particles and interactions) of the UV-completion.

Direct detection signatures of neutralino DM have been discussed extensively in the literature. However, no systematic study exists for DM candidates in strongly coupled scenarios (with the exception of ref.~\cite{Bagnasco:1993st}, which focused on techni-baryon DM in Higgsless theories).

In this paper we present a model-independent analysis of the most relevant direct detection signatures of a {\emph{gauge-singlet}} DM candidate of arbitrary spin $S$. We focus on scenarios with WIMPs heavier than $\sim100$ GeV, for which direct detection experiments have the best sensitivity, and assume that the new physics follows Naive Dimensional Analysis. This formalism is especially suited for strongly coupled dark sectors, but our results have a much wider validity and can be easily generalized to models with light DM and light mediators.

The operator analysis is presented in section~\ref{sec:sig}. When restricted to real DM candidates, the effective Lagrangian includes the interactions discussed for neutralino DM in the MSSM (see for instance~\cite{Drees:1993bu}\cite{Jungman:1995df}), whereas for Dirac or complex scalars our results extend~\cite{Bagnasco:1993st} by including couplings to the Higgs and SM fermions. The current direct detection constraints are then analyzed using LUX~\cite{Akerib:2013tjd} and XENON100 (2012)~\cite{Aprile:2012nq} data in section~\ref{sec:Xe}.

Unnaturalness may not only be favored experimentally, but can also be theoretically attractive. In~\cite{ArkaniHamed:2004fb} it was emphasized that unnatural realizations of Supersymmetry can be viewed as elegant theories for charge quantization, dark matter, and (more recently) the Higgs mass~\cite{ArkaniHamed:2012gw}\cite{Arvanitaki:2012ps}. In the same spirit, in section~\ref{sec:theory} we will argue that unnatural strongly coupled models (warped extra dimensions) represent compelling theories for the Higgs mass, flavor, and dark matter.

\section{Effective theory for Dark Matter direct detection}
\label{sec:sig}

We assume that the dark matter (DM) is a SM singlet field $X$ of mass $m_X$, and focus on elastic processes $X(p)T(k)\to X(p')T(k')$, with $T$ a target nuclei.~\footnote{For the DM masses and the (weak) couplings we consider here, the DM will penetrate the atmosphere and scatter off target nuclei as an ordinary WIMP.}

The effective Lagrangian is defined at a scale $\Lambda\gtrsim v_w\approx246$ GeV, as the sum of a (infinite) number of local operators ${\cal O}_{\alpha}$:
\ba\label{Leff}
{\cal L}_{\rm eff}=\sum_\alpha{\cal O}_\alpha.
\ea
Here $\Lambda$ should be understood as the typical mass of the fields mediating the interactions between DM and the SM. We will comment in section~\ref{sec:con} how the analysis should be modified in the presence of light mediators.

Making use of the Fierz identities, any ${\cal O}_{\alpha}$ can be written as the product of a SM-singlet operator ${O}_{\alpha}^A$, built solely of SM fields, and a DM bilinear
\ba\label{Oeff}
{\cal O}_\alpha\propto\left({\rm DM~bilinear}\right)^A~{O}^A_{\alpha},
\ea
with $A$ denoting a Lorentz index. It follows that the S-matrix elements for DM-nuclei scattering can be split into a nuclear matrix element of $O_\alpha^A$ and a DM bilinear 
\ba\label{Oeff'}
\langle X(p'),{s'}|\left({\rm DM~bilinear}\right)^A|X(p),s\rangle\propto({\cal T}^A)_{s's}, 
\ea
with $s,s'=1,\cdots,2S+1$ spin indices. Rather than considering DM operators we will be dealing directly with the matrix elements. The utility of this formalism is that it considerably simplifies the study of WIMPs with arbitrary spin $S$, because the auxiliary fields that would otherwise be needed to describe relativistic DM operators are automatically decoupled in physical processes.

As usual, the SM operators of large dimensionality are suppressed by higher powers of the heavy scale and can be neglected. We cannot always apply the same logic to the DM bilinears, though. Indeed, the typical momentum flowing in the DM operator is of order the DM mass $m_X$, and this may well be of order $\Lambda$ or even larger. In complete generality, one should therefore organize the effective field theory expansion according to the dimension of the SM operators $O_\alpha^A$, which is what we will do next. In section~\ref{sec:bilinears} we will then identify the corresponding DM bilinears at leading order in a ``heavy DM expansion". 

The result of this program is summarized in table~\ref{table}. In the first row are collected the most relevant SM operators, whereas ${\cal T}^A({\mathbb{C}})$ and ${\cal T}^A({\mathbb{R}})$ are the matrix elements for complex (${\mathbb{C}}$) and real (${\mathbb{R}}$) DM. Note that one can relate the transformation properties under P,T of the tensors ${\cal T}^A$ to those of the corresponding DM operator (using relations such as (\ref{CPT}) below). For example, time-reversal would forbid the electric dipole operator $B_{\mu\nu}v^\mu s^\nu$. Finally, $\left.{\cal M}_N\right|_{\rm SI}$ shows the functional dependence in the spin-independent nucleon-DM scattering amplitude; spin-dependent processes are discussed in section~\ref{sec:majo}.

An explicit connection between relativistic quantum field theory operators and matrix elements ${\cal T}^A$ can be given in a light DM scenario, $m_X\ll\Lambda$, in which case the usual power counting can be applied to the DM bilinears as well. The dominant operators and the corresponding matrix elements are shown for light $S=0,1/2,1$ DM in table~\ref{tableNR} (see Appendix~\ref{app:tensors} for details).


\begin{table}[t]
\begin{center}
\begin{tabular}{c|c|c|c|c} 
\rule{0pt}{1.2em}%
${O}^A_{\alpha}$ & $g'{{B}^{\mu\nu}}$  & ${H^\dagger i\overleftrightarrow{D}^\mu H}, \overline{q}\bar\sigma^\mu q$ & $g_s^2GG, \overline{q}Hq, {H^\dagger H}$ & $g_s^2G^{\{\mu}_\alpha G^{\nu\}\alpha}, \overline{q}\bar\sigma^{\{\mu} i{D}^{\nu\}}q$\\
\hline\hline
${\cal T}^A({\mathbb{C}})$ & ${\bf \varepsilon^{\mu\nu\alpha\beta}v_\alpha s_\beta, v^\mu s^\nu}$ & ${\bf v^\mu}$ & ${\bf 1}$ & ${\bf v^\mu v^\nu, s^\mu s^\nu, s^\mu v^\nu}$\\
$\left.{\cal M}_N\right|_{\rm SI}$ & $\frac{(\overrightarrow{v_\perp}\wedge\overrightarrow{s})\cdot \overrightarrow{iq}}{\overrightarrow{q}^2}, \frac{\overrightarrow{s}\cdot \overrightarrow{iq}}{\overrightarrow{q}^2}$ & $1$ & $1$ & $1,1,\overrightarrow{s}\cdot\overrightarrow{v_\perp}$\\
\hline
${\cal T}^A({\mathbb{R}})$ & ${\bf s^\mu iq^\nu}$ & ${\bf s^\mu}$ & ${\bf 1}$ & ${\bf v^\mu v^\nu, s^\mu s^\nu}$\\
$\left.{\cal M}_N\right|_{\rm SI}$ & $\overrightarrow{s}\cdot\overrightarrow{v_\perp}$ & $\overrightarrow{s}\cdot\overrightarrow{v_\perp}$ & $1$ & $1,1$
\end{tabular}
\end{center}
\caption{\small List of the most relevant SM operators $O_\alpha^A$ contributing to DM direct detection (first row). In the second and third rows we collect the leading tensor structures ${\cal T}^A$ of the DM bilinears for complex (${\mathbb{C}}$) and real (${\mathbb{R}}$) DM, respectively, as well as the functional form of the non-relativistic spin-independent nucleon-DM matrix element $\left.{\cal M}_N\right|_{\rm SI}$ (with $v_\perp=v-q/2\mu_N$). CP is not imposed. See text for details.
\label{table}}
\end{table}

\subsection{SM operator analysis}
\label{sec:SM}

We now study a basis for the most relevant SM-singlet operators. Under the assumption that the masses and couplings of the dark sector follow Naive Dimensional Analysis (NDA) expectations, the main discriminant is the scaling dimension $d_\alpha$ of the SM operator $O_\alpha^A$. We limit ourselves to $d_\alpha\leq4$.

We neglect interactions involving leptons because not directly relevant to DM-nuclei scattering. This remains true even in scenarios in which DM couplings to gauge bosons and quarks are suppressed, in which case the main effect of these operators would be to renormalize the couplings in table~\ref{table}.

There are two SM singlet operators with $d_\alpha=2$:
\ba\label{dim2}
B^{\mu\nu},~~~~~~~~~~~~~~~~~~H^\dagger H.
\ea
We do not consider the dual field $\widetilde{B}_{\mu\nu}$ as an independent operator.

The operators of dimension $d_\alpha=3$ are:
\ba\label{dim3}
H^\dagger i\overleftrightarrow{D}^\mu H,~~~~~~~~~~~~\partial_\mu(H^\dagger H),~~~~~~~~~~~~~\overline{q}\bar\sigma^\mu q,~~~~~~~~~~\partial_\alpha B^{\mu\nu},
\ea
where $q$ stands for any (electroweak singlet or doublet) SM quark and $A^\dagger\overleftrightarrow{D}_\mu B\equiv A^\dagger D_\mu B-(D_\mu A)^\dagger B$. Flavor-violating couplings are in principle allowed, but only contribute at loop-level in direct detection experiments and are very constrained by $\Delta F\neq0$ processes. We do not consider $\partial_\mu(H^\dagger H)$ because it is a momentum-suppressed version of $H^\dagger H$. For similar reasons we do not include in table~\ref{table} the various contractions of $\partial_\alpha B^{\mu\nu}, \partial_\alpha \widetilde B^{\mu\nu}$. 

Finally, at $d_\alpha=4$ we find:
\ba\label{dim4}
&&\overline{q}Hq~~~~~~\overline{q}\sigma^{\mu\nu}H q~~~~~~\overline{q}\bar\sigma^{\mu} {D}^{\nu} q~~~~~~G_{\mu\nu}^aG_{\alpha\beta}^a,
\\\no
&&B_{\mu\nu}B_{\alpha\beta}~~~~~~{\rm tr}(W_{\mu\nu}W_{\alpha\beta})~~~~~~\partial_\alpha\partial_\beta B_{\mu\nu}~~~~~~B_{\mu\nu}H^\dagger H~~~~~~{\rm tr}(H^\dagger W_{\mu\nu}H),\\\no
&&(D_\mu H)^\dagger D_\nu H~~~~~~\partial_\nu(H^\dagger D_\mu H)~~~~~~(H^\dagger H)^2,
\ea
where all possible contractions of the indices (also with the Levi-Civita tensor) are in principle possible. The list of dim-4 operators in table~\ref{table} does not include the operators in the last line of~(\ref{dim4}) because these can be reduced to lower dimensional operators after integration by parts (or removed by redefining the Higgs field), or simply do not contribute to 2-to-2 DM-nuclei scattering at leading order. Also, $B_{\mu\nu}H^\dagger H, {\rm tr}(H^\dagger W_{\mu\nu}H)$ just lead to small corrections to processes mediated by $B_{\mu\nu}$. Similarly, $(H^\dagger H)^2$ is subleading compared to $H^\dagger H$. As before, derivatives of $B_{\mu\nu}$ and $\widetilde B_{\mu\nu}$ are neglected. Furthermore, we ignored the squared of the field strength of the electroweak bosons because they lead to rates that are suppressed compared to those induced by $G_{\mu\nu}^aG_{\alpha\beta}^a$ (see for instance~\cite{Weiner:2012cb}\cite{Pospelov:2000bq}).

Some of the quark operators in~(\ref{dim4}) also appear to be ``subdominant" or redundant. For example, DM with anti-symmetric bilinears can couple to $\overline{q}\sigma^{\mu\nu}H q, \overline{q}\bar\sigma^{[\mu} {D}^{\nu]} q$, provided the dark sector has unsuppressed interactions with quarks. Yet, in these cases couplings to vector quark operators such as those in (\ref{dim3}), or $\overline{q}Hq$, are generally present as well, and dominate. As a result, the anti-symmetric SM tensors in the first line of (\ref{dim4}) have been excluded from table~\ref{table}.

We are thus left with the symmetric combinations of $\overline{q}\bar\sigma^{\mu} {D}^{\nu} q$, and $\overline{q}Hq$. The trace of $\overline{q}\bar\sigma^{\{\mu} {D}^{\nu\}} q$, i.e. the kinetic operator $\overline{q}\bar\sigma^{\mu} {D}_{\mu} q$, can be removed in favor of the Yukawa coupling $y_{\rm SM}\overline{q}Hq$ by making use of the equations of motion (or equivalently a field redefinition).~\footnote{Note that the equations of motion can be used even for quarks and gluons far off the mass-shell.} Therefore, only the symmetric-traceless, and hermitian, part
\ba\label{OXqq}
O_{Xqq}^{\mu\nu}\equiv\frac{i}{4}\left[\overline{q}\left(\gamma^{\mu}\overleftrightarrow{D}^{\nu}+\gamma^{\mu}\overleftrightarrow{D}^{\nu}\right) q-\frac{g^{\mu\nu}}{2}\overline{q}\gamma^{\mu}\overleftrightarrow{D}_{\mu} q\right],
\ea
will be of interest here. In particular, the combination $\partial^\nu(\overline{q}\gamma^\mu q)$ is momentum-suppressed, and can be reduced to a dimension-3 coupling. The analogous axial operators are obtained by replacing $\gamma^\mu\to\gamma^\mu\gamma^5$ in~(\ref{OXqq}). In this latter case one should note that the flavor-singlet combination $\partial_\mu(\overline{q}\gamma^\mu\gamma^5 q)$ mixes with $G\widetilde G$, so it cannot be neglected in general.

The most important combinations of $G_{\mu\nu}^aG_{\alpha\beta}^a$ are the 2-index symmetric and trace parts:
\ba\label{GGGG}
O_{XGG}^{\mu\nu}\equiv-G^{\mu}_\alpha G^{\nu\alpha}+\frac{g^{\mu\nu}}{4}G_{\alpha\beta} G^{\alpha\beta}~~~~~~~~G_{\alpha\beta} G^{\alpha\beta}.
\ea
These contribute to the QCD energy-momentum tensor, so their nucleon matrix elements are $O(m_N^2)$. On the other hand, the matrix elements of all other combinations involve either the nucleon spin or the momentum transfer, and hence lead to suppressed rates. In particular, the couplings to $G_{\mu\nu}^a\widetilde G_{\alpha\beta}^a$ are expected to be comparable in magnitude to those of the CP-even tensors in a generic theory, but the rates are much smaller.

\subsubsection{Renormalization group and matching}
\label{RGeffects}

It is often necessary to evolve the operators from the heavy mass $\Lambda$, where ${\cal L}_{\rm eff}$ is formally defined, down to $\mu\ll\Lambda$, where the matrix elements of $O_\alpha^A$ are extracted. The dominant renormalization group (RG) effects are expected to come from QCD, so only quark and gluon operators will be considered.~\footnote{The nuclear matrix elements of these operators suffer from large hadronic uncertainties, which are often more important than the RG effects themselves.}

In the chiral limit $m_{u,d,s}=0$ the light quark vector operators in table~\ref{table} do not run. Those involving the heavy quarks ($q=c,b,t$) are matched onto gluon operators (see~\cite{Fan:2010gt} and references therein) and provide negligible corrections (in the generic scenarios considered here). We will therefore use the standard nucleon matrix elements:
\ba\label{matchV}
\langle T'|\sum_qc_q(\overline{q}\gamma^\mu q)(\Lambda)|T\rangle=\langle T'|(c_u+2c_d)(\overline{n}\gamma^\mu n)(\mu)+(2c_u+c_d)(\overline{p}\gamma^\mu p)(\mu)|T\rangle,
\ea
with $n,p$ the neutron and proton fields.

For the axial operators one can use similar arguments to discard the interactions involving heavy quarks. However, because of the axial anomaly any $\overline{q}\gamma^\mu\gamma^5q$ (where $q=u,d$ or $s$) has the same anomalous dimension $\gamma_s=16N_f\left(\frac{\alpha_s}{4\pi}\right)^2+O(\alpha_s^3)$ of the singlet axial current $\sum_{q=u,d,s}\overline{q}\gamma^\mu\gamma^5q$. Using the three-loop anomalous dimension $\gamma_s$~\cite{Larin:1993tq} and QCD beta function $\beta_s$, and for $\mu_N\sim m_N$, we find:
\ba\label{matchA}
\langle T'|\sum_qc_q(\overline{q}\gamma^\mu\gamma^5 q)(\Lambda) |T\rangle&=&\eta\,\langle T'|\sum_{N=p,n}\sum_{q=u,d,s}c_q\Delta^{(N)}_q(\overline{N}\gamma^\mu\gamma^5 N)(\mu_N)|T\rangle,\\\no
\eta&=&0.80\times\exp\left(+\frac{12}{7\pi}\alpha_s(\Lambda)+O(\alpha^2_s(\Lambda))\right)
\ea
where $\eta$ parametrizes the RG effects from $\Lambda>m_t$, whereas a list of the nucleon matrix elements $\Delta_q^{(N)}$s used in the literature is given in~\cite{DelNobile:2013sia}. Despite $\mu_N$ is a rather low scale, the three-loop contribution provides just a few percent correction to the leading (2-loop) order. The impact of the uncertainties on the quark masses is of the same magnitude. This suggests that the perturbative expansion is reliable.

The last five operators in table \ref{table}, as well as their axial versions, are renormalized already at one-loop (after the Higgs is integrated out, $H^\dagger H$ can be turned into $m_q\overline{q}q$). These operators may be divided into four classes (see section~\ref{sec:SM} for our conventions): the CP-even $(O_{XGG}^{\mu\nu}, O_{Xqq}^{\mu\nu})$, $(GG, m_q\overline{q}q)$, as well as their CP-odd counterparts. 

The matrix elements of the spin-2 operators are efficiently extracted by deep-inelastic data. In this case the measurement is more easily obtained at $\mu\gg1$ GeV, and the RG effects from $\Lambda$ down to $\mu_N$ are well described by a one-loop analysis (for the CP-even tensors this is a textbook result, whereas for the CP-odd see~\cite{Ahmed:1976ee}). $O_{Xqq, XGG}^{\mu\nu}$ enter the QCD stress-energy tensor, so:
\ba\label{2S}
\langle N|O_{Xqq, XGG}^{\mu\nu}(\mu)|N\rangle&=&2{f^{(N)}_{q2, G2}(\mu)}\left(k_\mu k_\nu-\frac{g_{\mu\nu}}{4}k^2\right)\\\no
f^{(N)}_{q2}(\mu)&=&\int_0^1 dx\, x(f^{(N)}_{q}(x,\mu)+f^{(N)}_{\bar q}(x,\mu)),\\\no
f^{(N)}_{G2}(\mu)&=&\int_0^1 dx\, x\,f^{(N)}_{G}(x,\mu),
\ea
where $f^{(N)}_{q, G}(x,\mu)$ are the PDFs of the corresponding quark/gluon renormalized at $\mu$.

On the other hand, the nucleon matrix elements for $GG, \overline{q}q$ are estimated via lattice QCD at $\mu_N\sim1$ GeV. Therefore, larger RG effects are expected. We will discuss the RG evolution in detail in Appendix~\ref{sec:RG}. We define~\footnote{Throughout the paper we use a relativistic normalization for the one-particle states.}
\ba\label{fTqdef}
f_{Tq}^{(N)}\equiv\frac{\langle N|m_q\overline{q}q|N\rangle}{2m_N^2}
\ea
and work at $O(\alpha_s^3)$ and leading order in a heavy quark expansion. The numerical result is:
\ba\label{result2'}
f_{Tt}^{(N)}&=&\frac{2}{27}\left(1.02-1.07\sum_{q=u,d,s}f_{Tq}^{(N)}\right)+O\left(\frac{1}{m_{c}^2}\right)\\\no
f_{Tb}^{(N)}&=&\frac{2}{27}\left(1.07-1.23\sum_{q=u,d,s}f_{Tq}^{(N)}\right)+O\left(\frac{1}{m_{c}^2}\right)\\\no
f_{Tc}^{(N)}&=&\frac{2}{27}\left(1.18-(1.47-1.49)\sum_{q=u,d,s}f_{Tq}^{(N)}\right)+O\left(\frac{1}{m_{c}^2}\right),
\ea
and
\ba\label{result3'}
\frac{\langle N|(g_s^2GG)(\Lambda)|N\rangle}{2m_N^2}&\equiv&-\frac{32\pi^2}{9}f_{TG}^{(N)}(\Lambda)+O\left(\frac{g^2_s(\Lambda)}{16\pi^2}\right)\\\no
&=&-\frac{32\pi^2}{9}\left(0.97-0.93\sum_{q=u,d,s}f_{Tq}^{(N)}\right)+O\left(\frac{g^2_s(\Lambda)}{16\pi^2},\frac{1}{m_{c}^2}\right).
\ea
The NLO corrections to the one-loop analysis of~\cite{Shifman:1978zn} are of order $10-20\%$, and in practice are relevant only when the DM coupling to the charm quark are large. The terms $O\left({1}/{m_{c}^2}\right)$ are estimated to be of the order a few percent. For more details see the appendix.

The RG evolution of $G\widetilde G, \sum_q\partial_\mu(\overline{q}\gamma^\mu\gamma^5q)$ can be found up to three-loops in \cite{Larin:1993tq}.

\subsection{DM matrix elements}
\label{sec:bilinears}

Direct detection processes have two small parameters:
\ba
\frac{|\overrightarrow{q}|}{m_X}\ll 1,~~~~~~~~~~~|{\overrightarrow{v}}|\ll1,
\ea
with $\overrightarrow{v}$ the velocity of the incoming DM particle in the lab frame, of order $\sim10^{-3}$, and $\overrightarrow{q}=\overrightarrow{p}-\overrightarrow{p}'$ the momentum transferred to the target, $|\overrightarrow{q}|\sim\mu_T |\overrightarrow{v}|$ ($\mu_T$ is the reduced mass of the DM-target system). The first hierarchy suggests to perform an expansion in momentum transfer, i.e. a heavy ``quark" effective field theory approach~\cite{Georgi:1990um}\cite{Jenkins:1990jv}. This was done in~\cite{Hill:2011be} for real DM with electroweak charges. The second indicates that we are dealing with a non-relativistic process. Refs.~\cite{Fan:2010gt}\cite{Fitzpatrick:2012ix} used this to perform a non-relativistic analysis of DM-nucleon scattering. Here we will instead fully exploit the Lorentz symmetry, which implies correlations among the different $SO(3)$ components of the Lorentz tensors.

The matrix elements $({\cal T}^A)_{s's}$ of the DM bilinears are first identified in the ``CM frame of the DM system", defined by $\overrightarrow{p}+\overrightarrow{p}'=0$, and then boosted to the lab frame (the little group rotation is the identity plus $O(v^2)$ corrections). The resulting matrix elements can be written, in momentum space, as a combination of the DM 4-velocity, the spin, and the 4-momentum transferred to the target, or equivalently:
\ba\label{4v}
v^\mu=\frac{(p+p')^\mu}{2m_X}~~~~~~~~~(s^\mu)_{s's}~~~~~~~~~q^\mu=(p-p')^\mu,
\ea
with $s,s'=1,\cdots,(2S+1)$. These satisfy $q^\mu v_\mu=s^\mu v_\mu=0$ and $v_\mu v^\mu=1-q^2/4m_X^2$, $s^\mu s_\mu=-S(S+1)$, with $S$ the DM spin. In writing the DM bilinears it is sometimes useful to note that $v^\mu, s^\mu, iq^\mu$ are hermitian operators, as emphasized in~\cite{Fitzpatrick:2012ix}.

The building blocks in~(\ref{4v}) are the same appearing in a heavy quark expansion~\cite{Georgi:1990um}\cite{Jenkins:1990jv}. Ultimately, the reason is that a massive 1-particle state is described by its 4-momentum and the Pauli-Lubanski vector, whereas soft DM scattering can be accounted for by local perturbations $O(q/m_X)$ around the free particle configuration.

\begin{table}[t]
\begin{center}
\begin{tabular}{|c|cccc|} 
\hline
$S=0$ & ${\cal T}_0$  & ${\cal T}_1^\mu$ & ${\cal T}_{2A}^{\mu\nu}$ & ${\cal T}_{2S}^{\mu\nu}$\\
\hline
${\mathbb{C}}$ & $\overline{X}X$ & $\overline{X}i\overleftrightarrow{\partial^\mu}X$ & $i\partial^{[\mu}\overline{X}\partial^{\nu]} X$ & $\partial^{\{\mu}\overline{X}\partial^{\nu\}}X$\\
\quad & $1$ & $2m_Xv^\mu$ & $m_Xv^{[\mu}iq^{\nu]}$ & $2m_X^2 v^{\{\mu}v^{\nu\}}$  \\
\hline
${\mathbb{R}}$ & ${X}X$ & $\partial^\mu({X}X)$ & --- & $\partial^{\{\mu}{X}\partial^{\nu\}} X$\\
\quad & $2$ & $-2iq^\mu$ & --- & $4m_X^2 v^{\{\mu}v^{\nu\}}$\\
\hline\hline
\rule{0pt}{1.2em}%
\rule{0pt}{1.2em}%
$S=1/2$ & ${\cal T}_0$  & ${\cal T}_1^\mu$ & ${\cal T}_{2A}^{\mu\nu}$ & ${\cal T}_{2S}^{\mu\nu}$\\
\hline
${\mathbb{C}}$ & $\overline{X}X$ & $\overline{X}{\gamma^\mu}X$ & $\overline{X}\sigma^{\mu\nu} X,~\overline{X}i\sigma^{\mu\nu}\gamma^5 X$ & $\overline{X}\gamma^{\{\mu} i\partial^{\nu\}} X$\\
\quad & $2m_X$ & $2m_Xv^\mu$ & $-4m_X\varepsilon^{\mu\nu\alpha\beta}v_\alpha s_\beta,~-4m_Xv^{[\mu}s^{\nu]}$ & $4m_X^2 v^{\{\mu}v^{\nu\}}$  \\
\hline
${\mathbb{R}}$ & $\overline{X}X$ & $\overline{X}{\gamma^\mu}\gamma^5X$ & $\overline{X}\gamma^{[\mu} i\partial^{\nu]} X,~\overline{X}\gamma^{[\mu}\gamma^5\partial^{\nu]} X$ & $\overline{X}\gamma^{\{\mu} i\partial^{\nu\}} X$\\
\quad & $4m_X$ & $8m_Xs^\mu$ & $-4m_X\varepsilon^{\mu\nu\alpha\beta}iq_{\alpha}s_{\beta},~2m_Xs^{[\mu}iq^{\nu]}$ & $8m_X^2 v^{\{\mu}v^{\nu\}}$\\
\hline\hline
\rule{0pt}{1.2em}%
$S=1$ & ${\cal T}_0$  & ${\cal T}_1^\mu$ & ${\cal T}_{2A}^{\mu\nu}$ & ${\cal T}_{2S}^{\mu\nu}$\\
\hline
${\mathbb{C}}$ & $\overline{X_\alpha}X^\alpha$ & $\overline{X_\alpha}i\overleftrightarrow{\partial^\mu}X^\alpha$ & $i\overline{X^{[\mu}}X^{\nu]},~i\varepsilon^{\mu\nu\alpha\beta}\overline{X_\alpha}X_\beta$ & $\overline{X^{\{\mu}}X^{\nu\}}$\\
\quad & $-1$ & $-2m_Xv^\mu$ & $-\varepsilon^{\mu\nu\alpha\beta}v_\alpha s_\beta,~v^{[\mu}s^{\nu]}$ & $v^{\{\mu}v^{\nu\}}-s^{\{\mu}s^{\nu\}}$ \\
\hline
${\mathbb{R}}$ & ${X_\alpha}X^\alpha$ & $\varepsilon^{\mu\alpha\beta\gamma}{X_\alpha}\partial_\beta X_\gamma$ & $\partial_\alpha{X^{[\mu}}\partial^{\nu]} X^\alpha$ & ${X^{\{\mu}}X^{\nu\}}$\\
\quad & $-2$ & $-2m_Xs^\mu$ & $m_X\varepsilon^{\mu\nu\alpha\beta}iq_\alpha s_\beta$ & $2(v^{\{\mu}v^{\nu\}}-s^{\{\mu}s^{\nu\}})$\\
\hline
\end{tabular}
\end{center}
\caption{\small Here we list,  for $S=0,1/2,1$ and a given Lorentz structure (singlet, vector, 2-index anti-symmetric and 2-index symmetric), the most relevant DM bilinears contributing in direct detection experiments (CP is not assumed). Below each operator we show the expression of the matrix element for Complex (${\mathbb{C}}$) and Real (${\mathbb{R}}$) DM at leading order in $q/m_X$. We defined $A^{[\mu}B^{\nu]}=A^\mu B^\nu-A^\nu B^\mu$, and similarly $A^{\{\mu}B^{\nu\}}=A^\mu B^\nu+A^\nu B^\mu$. More operators and details can be found in the text and in Appendix~\ref{app:tensors}.
\label{tableNR}}
\end{table}

\subsubsection{Lorentz-singlet}

The only Lorentz scalars we can build with~(\ref{4v}) are $m_X^2,s^\mu iq_\mu$ and $q^2$. The dominant DM bilinears are therefore a T-even and T-odd scalar, whose matrix elements simply read
\ba\label{T0}
{\cal T}_0&=&1+O(q^2/m_X^2),\\\no
&&s^\mu iq_\mu+O(q^2/m_X^2),
\ea
where spin-indices ${s,s'}$ are understood.

\subsubsection{Vector and axial-vector}

At zeroth order in $q$, the DM vectors can be written as:
\ba\label{T1}
{\cal T}^\mu_1&=&v^\mu+O(q/m_X)~~~~({\rm Complex~DM})\\\no
&&s^\mu+O(q/m_X)~~~~({\rm Real~\&~Complex~DM}).
\ea
Matrix elements involving higher powers of the spin operator, e.g. $s^\alpha s^\mu s_\alpha$, can be reduced to~(\ref{T1}) by repeated use of the $SU(2)$ algebra.

Now, complex DM admits both matrix elements ${\cal T}^\mu_{1}=av^\mu+bs^\mu$, in which case the main signature is typically induced by the vector coupling $\propto v^\mu$. Self-conjugate (real) DM does not carry a continuous charge, so these particles couple to~(\ref{dim3}) via axial currents $\propto s^\mu$ (i.e. $a=0$). One way to see this is as follows. By CPT invariance, the matrix element of the DM bilinear between two DM states, ${\cal T}^A$, is related to the matrix element between CPT-transformed states, $({\cal T}^A)^c$, via the relation 
\ba\label{CPT}
{\cal T}^A=\eta({\cal T}^A)^c, 
\ea
with $\eta$ defined by CPT$($DM bilinear$)$CPT$^{-1}=\eta($DM bilinear$)$. Since the DM operator is in this case a Lorentz vector, we can choose $\eta=-1$ in any unitary relativistic theory. Real DM in addition satisfies $({\cal T}^\mu_{1})^c=av^\mu-bs^\mu$, since a CPT transformation on the 1-particle state changes its spin but leaves the momentum unchanged. It follows that the component $\propto v^\mu$ must vanish.

Including corrections of order $q$, four additional Lorentz vectors can be written down up to $O(q/m_X)$, namely $iq^\mu, v^\mu(s^\alpha iq_\alpha), \varepsilon^{\mu\alpha\beta\gamma}v_\alpha s_\beta iq_\gamma, s^\mu(s^\alpha iq_\alpha)$. These are small compared to~(\ref{T1}), and can always be neglected except when the DM is a real scalar. The only possible candidate in this latter case is $iq^\mu+O(q^2/m_X^2)$, which results in very suppressed scattering rates.

\subsubsection{Anti-symmetric bilinears}

Neglecting $q^\mu$, the only Lorentz covariant 2-index antisymmetric tensors are  
\ba\label{TA}
{\cal T}^{\mu\nu}_{2A}&=&\varepsilon^{\mu\nu\alpha\beta} v_\alpha s_\beta+O(q/m_X)~~~~~~({\rm Complex~DM}),\\\no
&&v^{[\mu} s^{\nu]}+O(q/m_X)~~~~~~~~~~~~({\rm Complex~DM}). 
\ea
To see this, note that in the frame $\overrightarrow{p}+\overrightarrow{p}'=0$, and neglecting $O(q/m_X)$ terms, the only non-trivial candidate for the ${ij}$ component is $\varepsilon^{ijk}s^k$, which can be uplifted to $\varepsilon^{\mu\nu\alpha\beta} v_\alpha s_\beta$. Contractions of this latter with the Levi-Civita tensor then gives us $v^{[\mu} s^{\nu]}$. Had we started looking at the $0i$ component, rather than $ij$, we would have arrived at the same result~(\ref{TA}).

By an argument similar to the one discussed for ${\cal T}_1^\mu$ we can prove that, in addition to have $S\neq0$, any DM field with (\ref{TA}) must be complex. Indeed, the DM operators must now be CPT invariant, so that eq.~(\ref{CPT}) holds with $\eta=1$. However, real DM has $(v^\mu s^{\nu})^c=-v^\mu s^\nu$, which is incompatible with~(\ref{CPT}). As a consequence, real DM has vanishing couplings to all SM 2-index antisymmetric tensors at leading order in $q$. This is a generalization of the statement that Majorana particles (or real vectors and scalars) do not have electromagnetic dipole moments.

We find six new tensors at next to leading order. The {{hermitian}} expressions can be written as $v^{[\mu} iq^{\nu]}$, $s^{[\mu} iq^{\nu]}$, $s^{[\mu} v^{\nu]}s^\alpha iq_\alpha$, and analogous contractions with $\varepsilon^{\mu\nu\alpha\beta}$. For complex DM, these result in small corrections to~(\ref{TA}), but for real DM they cannot be neglected. By studying their transformations under CPT we single out 
\ba
{\cal T}^{\mu\nu}_{2A}&=&s^{[\mu} iq^{\nu]}+O(q^2/m^2_X)~~~~~~~~~~~~~({\rm Real~DM}),\\\no
&&\varepsilon^{\mu\nu\alpha\beta}s_\alpha iq_\beta+O(q^2/m^2_X)~~~~~~~({\rm Real~DM})
\ea
as the only CPT-even combinations, and therefore the dominant anti-symmetric bilinears for real DM. By formally performing an ``integration by parts", these can be reduced to couplings between an axial ${\cal T}_1^\mu$ and the derivative of the (anti-symmetric) SM operator. (Note that $\varepsilon^{\mu\nu\alpha\beta}s_\alpha iq_\beta B_{\mu\nu}=0$ by the Bianchi identity.) Finally, real scalars do not have antisymmetric bilinears.

\subsubsection{Symmetric-traceless bilinears}

The dominant 2-index symmetric bilinears can be written as:
\ba\label{TS}
{\cal T}_{2S}^{\mu\nu}&=&\left(v^\mu v^\nu-\frac{g^{\mu\nu}}{4}\right)+O(q/m_X),\\\no
&&\left(s^\mu s^\nu+s^\nu s^\mu+S(S+1)\frac{g^{\mu\nu}}{2}\right)+O(q/m_X),\\\no
&&\,\left(s^\mu v^\nu+s^\nu v^\mu\right)+O(q/m_X)\ea
For spin-1/2 DM, the $s^\mu s^\nu$ structure is equivalent (modulo a factor $1/2$) to the first one in~(\ref{TS}) because of the anti-commutation properties of the Pauli matrices, while is clearly trivial for scalar DM. It first appears as a new coupling for $S=1$, as we find in table~\ref{tableNR}. The structure $v^\mu s^\nu$ is absent in the case of real DM. Because any DM particle admits at least the first coupling in~(\ref{TS}), the corrections $O(q/m_X)$ will be ignored.

\section{Analysis of the LUX and XENON100 bounds}
\label{sec:Xe}

In this section we will present the current direct detection (DD) bounds on the couplings of table~\ref{table}, assuming that $X$ constitutes the totality of DM with local density of $\rho_X=0.3$ GeV$/$cm$^3$, and turning on the operators ${\cal O}_\alpha$ in~(\ref{Leff}) one at a time. We will mostly focus on the results of LUX~\cite{Akerib:2013tjd} and XENON100 (2012)~\cite{Aprile:2012nq}, which provide the strongest constraints.

At leading order in the DM-SM couplings, the DM-nuclei amplitude can be written as: 
\ba\label{amplitude}
\sum_\alpha\left.{\cal M}_T\right|_\alpha&\equiv&\sum_\alpha\langle X'T'|{\cal{O}}_{\alpha}|XT\rangle\\\no
&=&\sum_\alpha {C^W_\alpha(\Lambda)}~({\cal T}^A)_{s's}~\langle T'|{O}^A_{\alpha}(\Lambda)|T\rangle,
\ea
where the ${\cal T}^A$s have been presented in section~\ref{sec:bilinears}, and the nuclear matrix elements may be extracted from the DM-nucleon matching conditions discussed in section~\ref{RGeffects}. 

All model-dependence is hidden in the Wilson coefficients $C^W_\alpha$, which we conventionally write according to Naive Dimensional Analysis (NDA): 
\ba
C^W_\alpha(\Lambda)=\#\, c_\alpha\, {m_X}{\Lambda}\left(\frac{1}{\Lambda}\right)^{n_D}\left(\frac{1}{f\sqrt{\Lambda}}\right)^{n_q}\left(\frac{1}{f}\right)^{n_H}.
\ea
We denote by $\Lambda$ and $\Lambda/f$ respectively the mass scale of the DM dynamics and the typical coupling among resonances, $n_{D,q,H}$ refer to the powers of derivatives ($[D_\mu,D_\nu]\sim gF_{\mu\nu}$), quark and Higgs fields, whereas $c_\alpha$ are numbers of order unity. We also introduce numerical factors $\#$ in order to exactly match the matrix elements of a Dirac fermion (see table~\ref{tableNR}).~\footnote{The overall $m_X$ ensures that the estimates are compatible with a large $N$ expansion for a heavy baryon DM~\cite{Witten:1979kh}. It also agrees with NDA expectations in the case of {\emph{light}} DM of spin $S=0,1/2,1$, provided a suppression factor $m_X/\Lambda$ is introduced in front of the ``chirality-violating" operators.}

The DD bounds are derived in two independent ways. We use the Mathematica code provided in~\cite{DelNobile:2013sia} for spin $S=1/2$ DM, as well as to produce the ``thin curves" in our figures. The maximally allowed number of events taken in~\cite{DelNobile:2013sia} is rescaled down so that at $m_X=1$ TeV the $90\%$ CL bound on the nucleon-DM cross section reads $1.2\times10^{-44}$ cm$^2$ for LUX and $2.0\times10^{-44}$ cm$^2$ for XENON100, in agreement with~\cite{Akerib:2013tjd} and~\cite{Aprile:2012nq}.

The analysis is independently checked, and generalized to arbitrary spin $S$, employing the following simplified procedure, valid for $m_X\gtrsim100$ GeV (see thick curves in the plots). We define the number of scattering events per nuclear recoil energy $E_R$ and unit time as
\ba\label{diffRate}
\frac{dR}{dE_R}&=&\frac{\rho_X}{m_X}N_T\int_{|\overrightarrow{v}|\,\geq\,v_{\rm min}(E_R)} d^3v~vf(\overrightarrow{v})~\frac{d\sigma_{T}}{dE_R}\theta(v_{\rm esc}-|\overrightarrow{v}+\overrightarrow{v}_{\rm e}|),
\ea
with $N_T={\rm detector~mass}/m_T$  the number of target nuclei (of mass $m_T$), which for both LUX and XENON100 we assumed to be a single species with $(A,Z)=(131,54)$ for simplicity. The minimal DM velocity required to transfer  a kinetic energy $E_R$ to the target is $v_{\rm min}(E_R)=\sqrt{m_TE_R/2\mu_{T}^2}$. We will often use $E_R^{\rm max}=2\mu_{T}^2v^2/m_T$.

$\sigma_{T}$ and $f(\overrightarrow{v})$ are, respectively, the cross section for scattering of DM and target nuclei and the DM velocity distribution, both defined in the lab frame. In the non-relativistic limit, the differential cross section reads
\ba
\frac{d\sigma_{T}}{dE_R}&=&\frac{\langle|{\cal M}_T|^2\rangle}{32\pi m_Tm_X^2v^2},
\ea
with $\langle|{\cal M}_T|^2\rangle$ the spin-averaged S-matrix element squared for DM-target scattering in eq.~(\ref{amplitude}). This involves a form factor $F(E_R)$. In the case of spin-independent scattering we take
\ba
F(E_R)&=&\frac{3 j_1(q r_0)}{qr_0}e^{-(q s)^2/2},
\ea
where $r_0^2\approx(1.44~A^{2/3}-5)$ fm$^2$ and $s\approx0.9$ fm.

The DM is assumed to follow a Boltzmann distribution with dispersion $v_0=220$ Km/s in the galactic frame. The earth velocity was taken to be $v_{\rm e}=[232+15\cos(2\pi (y-y_0))]$ Km/s with $y_0=152/365$. We used an escape velocity from the galactic center of $v_{\rm esc}=544$ Km/s. These assumptions, together with $\rho_X$, are all plagued by large uncertainties.

We convolve the differential rate~(\ref{diffRate}) with the (energy-dependent) efficiencies reported by the collaborations, use $N_T=118 (34)$ Kg/A$m_p$, and then integrate over $E_R\geq3~{\rm keV}~(E_R\in[6.6~{\rm keV},43.3~{\rm keV}])$ and a whole year for LUX (XENON100). To account for the smaller exposure we multiply by a factor of $85.3$ days/yr ($224.6$ days/yr). The $90\%$ CL bound on the standard spin-independent nucleon-DM cross section matches the LUX (XENON100) value of $1.2\times10^{-44}$ cm$^2$ ($2\times10^{-44}$ cm$^2$) -- at $m_X=1$ TeV -- after requiring that the thus obtained number of events be smaller than $5.3$ ($5.7$).

This simplified analysis does not include the acceptance in the conversion of the scintillation photons into photoelectrons, scintillation efficiencies, etc. These latter effects become less and less important as the DM mass increases.  Consistently, the results are in very good agreement with those obtained with the Mathematica code of~\cite{DelNobile:2013sia} for $m_X\gtrsim100$ GeV.

\subsection{DM dipole moments}
\label{sec:dipoles}

We begin our analysis with the couplings to $B_{\mu\nu}$ of complex DM with spin. Following table~\ref{table} and our conventions, we define:
\ba\label{Om}
\left.{\cal M}_T\right|_{Xm}&\equiv&\langle X'T'|{\cal O}_{Xm}|XT\rangle\\\no
&=&c_{Xm}\frac{4m_X}{\Lambda}~ {\bf \left(\varepsilon^{\mu\nu\alpha\beta}v_\alpha s_\beta \right)}~ \langle T'|g'B_{\mu\nu}|T\rangle.
\ea
In the non-relativistic limit the operator ${\cal O}_{Xm}$ describes the interaction $(2m_X)\mu~\overrightarrow{s}\cdot\overrightarrow{B}$, with $\overrightarrow{B}$ the magnetic field and $\mu=4ec_{Xm}/\Lambda$ the DM magnetic dipole moment.~\footnote{The NDA expectation $c_{Xm}\sim1$ is compatible with the lattice results of~\cite{Appelquist:2013ms}.}

As explained above, the numerical factors in~(\ref{Om}) are chosen so that the resulting matrix elements agree with those of $c_{Xm}\overline{X}\sigma^{\mu\nu}Xg'B_{\mu\nu}/\Lambda$, with $X$ a fermion. Table~\ref{tableNR} tells us that the very same bound on $\Lambda/c_{Xm}$ would be obtained if we considered, for example, vector DM with $4c_{Xm}m_Xi\overline{X}^{[\mu}X^{\nu]}g'B_{\mu\nu}/\Lambda$. Similar considerations hold for all the interactions discussed below.

The operator ${\cal O}_{Xm}$ mediates both a spin-independent and a spin-dependent interaction. The latter is a very small effect and will be neglected. The spin-independent part of the differential rate reads
\ba\label{Xm}
\left.\frac{d\sigma_T}{dE_R}\right|_{\rm SI}=\frac{4}{3}S(S+1)\frac{(e^2Z)^2}{\pi}\left|\frac{c_{Xm}}{\Lambda}\right|^2\frac{|F(E_R)|^2}{E_R}\left(1-\frac{E_R}{E_R^{\rm max}}\right),
\ea
with $E_R=\overrightarrow{q}^2/2m_T$ the kinetic energy transferred to the target of electric charge $Ze$. In deriving~(\ref{Xm}) we used
\ba
{\rm tr}(s^is^j)=\frac{\delta^{ij}}{3}{\rm tr}(\overrightarrow{s}\cdot\overrightarrow{s})=\frac{\delta^{ij}}{3}S(S+1)(2S+1),
\ea
and then averaged over the incoming spins. 

Using~(\ref{Xm}), the 90\% CL bound from LUX is
\ba\label{LUXm}
\Lambda\gtrsim150~{\rm TeV}~|c_{Xm}|\left(\frac{4}{3}S(S+1)\right)^{1/2}\sqrt{\frac{1~{\rm TeV}}{m_X}}. 
\ea
For $m_X=\Lambda$, $c_{Xm}=1$, $S=1/2$, the bound becomes $m_X\gtrsim28$ TeV. Comparing to the corresponding result from XENON100, that approximately reads as in (\ref{LUXm}) except for the replacement $150\to93$, we see that LUX appreciably benefits from the lower energy threshold.

As anticipated in our discussion around eq.~(\ref{TA}), the rate~(\ref{Xm}) is valid for any spin $S$, whereas the $O(q/m_Xv)=O(\mu_T/m_X)$ corrections (note that the leading term is $O(v)$) are model-dependent. Indeed, our expression agrees at leading order with the differential rate shown in~\cite{Banks:2010eh}, obtained for {\emph{light}} $S=1/2$ DM.~\footnote{In this respect, we should mention that we do not agree with the claim in~\cite{Bagnasco:1993st} and~\cite{Barger:2010gv} that the formula for light $S=1/2$ DM ($m_X\ll\Lambda$) generalizes to arbitrary spin modulo a simple rescaling of a factor of $\frac{4}{3}S(S+1)$. That the subleading corrections $O(\mu_T/m_X)$ are model dependent can be seen by noting, for example, that in the case of fermion DM additional $O(\mu_T/m_X)$ terms arise from operators such as $\overline{X}\gamma^{[\mu}\partial^{\nu]} X$, which cannot be neglected in general. For $S\geq1$ additional spin structures also exist at that order. See Appendix~\ref{app:tensors}.} In figure~\ref{dipoles} (left) we show the LUX, XENON100, and CDMS-Ge $90\%$ CL bounds in this latter case, including the spin-dependent contribution. For light DM the systematic uncertainties of the experiment become relevant, but the result quickly reduces to (\ref{LUXm}), $\Lambda\gtrsim93~{\rm TeV} |c_{Xm}|\sqrt{{1~{\rm TeV}}/{m_X}}$, and $\Lambda\gtrsim19~{\rm TeV} |c_{Xm}|\sqrt{{1~{\rm TeV}}/{m_X}}$ as soon as $m_X$ is above $m_T\sim130$ GeV. Ref.~\cite{Frandsen:2013bfa} recently derived similar bounds for $S=1/2$.

\begin{figure}
\begin{center}
\includegraphics[width=2.6in]{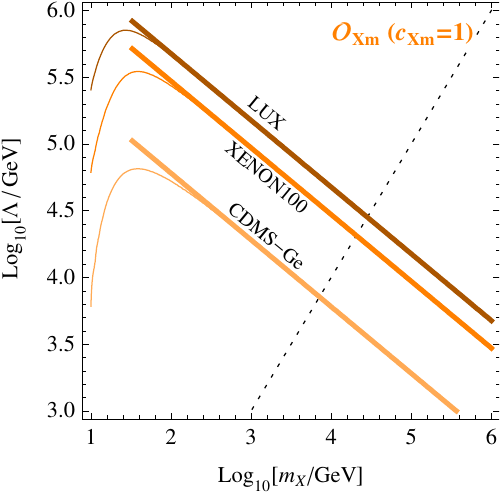}~~~~~~~~\includegraphics[width=2.6in]{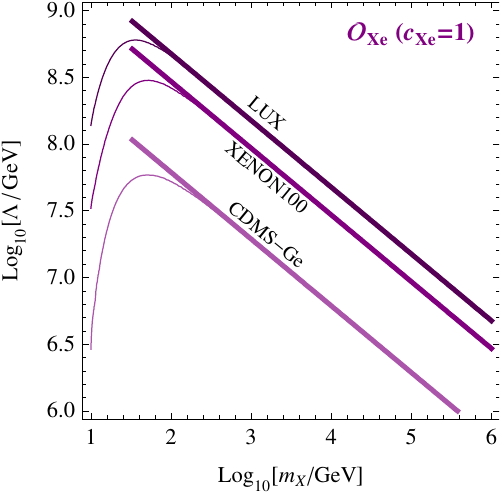}
\caption{\small $90\%$ CL bounds from LUX, XENON100(2012), and CDMS-Ge on the scale $\Lambda$ suppressing the magnetic (left) and electric (right) DM dipole moments for $S=1/2$. The thin curves are found with the Mathematica code provided in~\cite{DelNobile:2013sia}, whereas the thick lines are obtained as explained at the beginning of section~\ref{sec:Xe}. These latter generalize to arbitrary DM spin $S$ as shown in the text. The dotted line refers to $\Lambda=m_X$. 
\label{dipoles}}
\end{center}
\end{figure}

We can go through a completely analogous discussion for the electric dipole operator, which we define as 
\ba\label{XeO}
\left.{\cal M}_T\right|_{Xe}=c_{Xe}\frac{8m_X}{\Lambda}~{\bf\left(v^\mu s^\nu\right)}~ \langle T'|g'B_{\mu\nu}|T\rangle.
\ea 
The non-relativistic limit is the familiar $(2m_X)d~\overrightarrow{s}\cdot\overrightarrow{E}$, with $d=4ec_{Xe}/\Lambda$. We find: 
\ba\label{Xe}
\left.\frac{d\sigma_T}{dE_R}\right|_{\rm SI}=\frac{4}{3}S(S+1)\frac{(e^2Z)^2}{\pi}\left|\frac{c_{Xe}}{\Lambda}\right|^2\frac{|F(E_R)|^2}{E_Rv^2},
\ea
where $|\overrightarrow{v}|=v$ and we again neglected a small spin-dependent contribution. The expression~(\ref{Xe}) coincides with that of~\cite{Bagnasco:1993st}\cite{Banks:2010eh}\cite{Barger:2010gv}. The (model-dependent) corrections to ${\cal O}_{Xe}$ are here truly $O(q/m_X)$, and hence always much smaller than one. The LUX, XENON100, and CDMS-Ge bounds are shown for $S=1/2$ in purple in the right panel of figure~\ref{dipoles}.

At DM masses above $\sim100$ GeV, the LUX constraint reads, for arbitrary $S$,
\ba\label{Xeb}
\Lambda\gtrsim150\times10^3~{\rm TeV}~|c_{Xe}|\left(\frac{4}{3}S(S+1)\right)^{1/2}\sqrt{\frac{1~{\rm TeV}}{m_X}}.
\ea
XENON100 gives an analogous bound, with $150$ replaced by $92$. The rate is now enhanced by $O(1/v^2)$ compared to the magnetic dipole. The constraint on the new physics scale is consistently a factor $\langle v^{-1}\rangle\sim10^3$ stronger. 

Eq. (\ref{Xeb}) represents a significant constraint, not only for strongly coupled dynamics but also in weakly coupled theories, where the electric-dipole operator first arises at loop level. For example, for $c_{Xe}\sim1/16\pi^2$ and $m_X=\Lambda$ the bound becomes $m_X\gtrsim97$ TeV, while even for $c_{Xe}\sim(1/16\pi^2)^2$ one finds $\Lambda\sim m_X\gtrsim3$.

\subsection{``Charge radius" operator}
\label{sec:gauge}

We now discuss the ``charge radius" operator:
\ba\label{XB}
\left.{\cal M}_T\right|_{XB}=c_{XB}\frac{2m_X}{\Lambda^2}{\bf\left(v^\mu\right)}~\langle T'|g'\partial_\nu B^{\mu\nu}|T\rangle.
\ea
As emphasized in section~\ref{sec:bilinears}, real DM does not admit this coupling, but it is expected to have an anapole operator $s_\mu \partial_\nu B^{\mu\nu}$, which will be analyzed in section~\ref{sec:majo}.

\begin{figure}
\begin{center}
\includegraphics[width=2.6in]{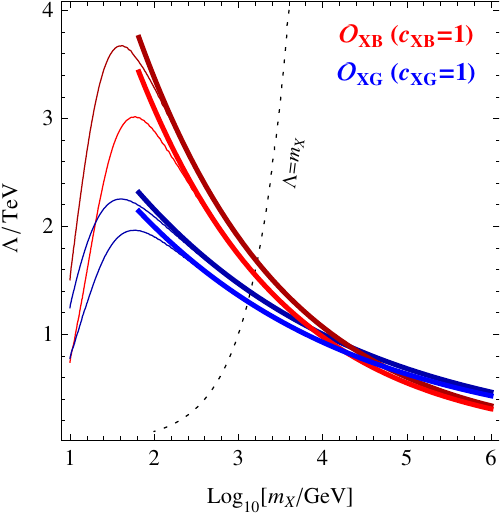}~~~~~~~~\includegraphics[width=2.75in]{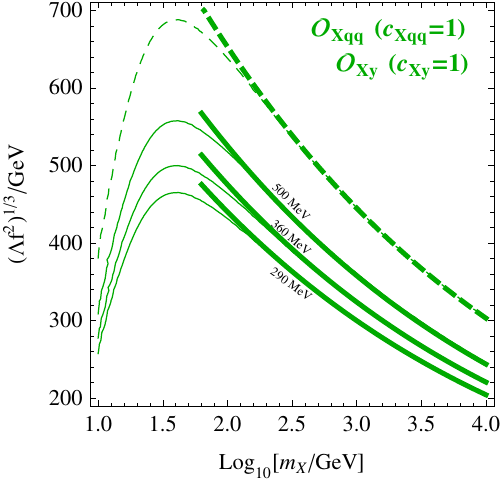}
\caption{\small Left: $90\%$ CL lower bounds (LUX and XENON100) on the scale $\Lambda$ suppressing the operators ${\cal O}_{XB}$ (red) and ${\cal O}_{XG}$ (blue) as a function of the DM mass, for took $f_{TG}^{(N)}=1$. LUX (mildly darker color) is always a bit more constraining. Right: bound on ${\cal O}_{Xy}$ (solid green) for $y_{ij}$ the SM Yukawa matrix and three values of the parameter $f_N=290, 360, 500$ MeV. In dashed green we show the constraint on ${\cal O}_{Xqq}$, assuming $\sum_qf^{(N)}_{q2}=1$. In the right panel, only the LUX constraints are presented; XENON100 is slightly weaker. The distinction between thin and thick lines is the same as in figure~\ref{dipoles}.
\label{figGauge}}
\end{center}
\end{figure}

At scales relevant for DM-nuclei scattering the effect of the $Z^0$ boson is suppressed by $q^2/m_Z^2$, and photon exchange dominates. The differential rate is given by
\ba
\left.\frac{d\sigma_T}{dE_R}\right|_{\rm SI}=\frac{\mu_{T}^2}{\pi}\left(c_{XB}\frac{e^2Z}{\Lambda^2}\right)^2\frac{|F(E_R)|^2}{E_R^{\rm max}}.
\ea
This result holds for complex DM of any mass and spin. The LUX and XENON100 bounds at $90\%$ CL are shown in the left of figure~\ref{figGauge} in solid red. For DM masses above $\sim m_T$ this depends on $m_X$ only via the local DM density, and the LUX bound becomes:
\ba\label{LXB}
\Lambda\gtrsim1.9\sqrt{|c_{XB}|}\left(\frac{\rm TeV}{m_X}\right)^{1/4}~{\rm TeV}.
\ea 
XENON100 provides a slightly weaker bound, of order $\Lambda\gtrsim1.7\sqrt{|c_{XB}|}({\rm TeV}/m_X)^{1/4}~{\rm TeV}$. At XENON1T one anticipates an improvement of a factor of $\sim100$ in  sensitivity over XENON100~\cite{XENON1T}, so the current $1.7$ TeV will become $\sim5.3$ TeV. The latter is comparable to that from the electroweak $S$-parameter, which roughly requires $m_Z^2/\Lambda^2\sim10^{-3}$.

\subsection{Gluon and ``quark mass" operators}
\label{sec:qg}

Let us now turn to the scalar couplings to gluons and quarks. The following results apply to complex and real DM of any spin.

First, consider
\ba\label{opXG}
\left.{\cal M}_T\right|_{XG}=c_{XG}\frac{2m_X}{\Lambda^3}{\bf \left(1\right)}\langle T'|g_s^2G^a_{\mu\nu}G^a_{\mu\nu}|T\rangle.
\ea
To estimate the effect of ${\cal O}_{XG}$ we employ the matching in (\ref{result3'}). Performing a coherent sum of the $N$s in the nuclei (as explained for instance in~\cite{Fitzpatrick:2012ix} and~\cite{Cirigliano:2012pq}), from~(\ref{result3'}) we obtain 
\ba
\left.\frac{d\sigma_T}{dE_R}\right|_{\rm SI}=\frac{\mu_{T}^2}{\pi}\left(c_{XG}f_{TG}^{(N)}\frac{32\pi^2m_NA}{9\Lambda^3}\right)^2\frac{|F(E_R)|^2}{E_R^{\rm max}}.
\ea
For $m_X>m_T$ LUX sets the following $90\%$ CL constraint:
\ba
\Lambda\gtrsim1.5~{\rm TeV}\times|c_{XG}f_{TG}^{(N)}|^{1/3}\left(\frac{\rm TeV}{m_X}\right)^{1/6}.
\ea
This decreases slower than (\ref{LXB}) with increasing DM mass, so the bound on ${\cal O}_{XG}$ becomes more important at larger masses. This is shown in the left of figure~\ref{figGauge} for $f_{TG}^{(N)}=1$ (the constraint on ${\cal O}_{XG}$ from LUX and XENON100 are both shown in blue).

Dark matter of any type can also couple to the SM quarks via
\ba\label{Oy}
\left.{\cal M}_T\right|_{Xy}=c_{Xy}\frac{2m_X}{\Lambda f^2}\,{\bf\left(1\right)}\,\langle T'|y_{ij}\overline{q_i}Hq_j|T\rangle,
\ea
where the matrix $y_{ij}$ is in general flavor-violating. After having replaced the Higgs with its vacuum expectation value we write $y_{ij}\overline{q_i}Hq_j\to {m_{ij}}\overline{q_i}q_j+O(h)$, where the $q$s are understood to be the mass eigenstates and $m_{ij}$ is hermitian. Because QCD conserves flavor, the effect of the off-diagonal terms in $m_{ij}$ are suppressed by a weak loop factor and powers of the Cabibbo angle. These are also very constrained by $\Delta F\neq0$ processes. We will therefore neglect the off-diagonal entries in what follows.

To estimate~(\ref{Oy}) we first derive an effective nucleon-DM coupling:
\ba\label{XXNN}
\langle N|y_{ij}\overline{q_i}Hq_j|N\rangle =2m_Nf_N=2m_N^2\sum_{q_i=u,d,s,c,b,t}f_{Tq_i}^{(N)}\frac{m_{ii}}{m_{q_i}}
\ea
where numerical values of $f_{Tc,Tb,Tt}^{(N)}$ as a function of $\sum_{q=u,d,s}f_{Tq}^{(N)}$ are given in eq.~(\ref{result2'}) at leading order in a heavy quark expansion. The rate reads:
\ba\label{rateOXy}
\left.\frac{d\sigma_T}{dE_R}\right|_{\rm SI}=c_{Xy}^2\frac{\mu_T^2}{\pi f^4}\frac{\left[Zf_p+(A-Z)f_n\right]^2}{\Lambda^2}\frac{|F(E_R)|^2}{E_R^{\rm max}}.
\ea

We will present our bounds for scenarios in which the $m_{ii}$s are the SM quark masses. In fact, we will see below that this is a defendable approximation in unnatural theories. In this class of models only the sum $\sum_{q=u,d,s}f_{Tq}^{(N)}$ appears. From chiral Lagrangian techniques the authors of~\cite{Gasser:1990ce} found $f_{Tu}^{(N)}\approx0.02$, $f_{Td}^{(N)}\approx0.03-0.04$, and $f_{Ts}^{(N)}\approx0.14$, but lattice data currently favor smaller values of $f_{Ts}^{(N)}$~\cite{fsLattice}. More recent analysis in chiral perturbation theory suggest that lower values of $f_{Ts}^{(N)}$ are in fact more plausible~\cite{Alarcon:2012nr}. An estimate of the small difference $f_{Tq}^{n}-f^p_{Tq}$ has been given recently in~\cite{Crivellin:2013ipa}, but will be neglected here. We use $\sum_{q=u,d,s}f_{Tq}^{(N)}=0.2, 0.4, 0.1$ (see also~\cite{Ellis:2008hf}\cite{Cheng:2012qr}), which translate into $f_N=\left[290,~~360,~~500\right]~{\rm MeV}$. The resulting $90\%$ CL bound on $\Lambda f^2$ from LUX are shown in the right of figure~\ref{figGauge} in solid green, and above $m_X\sim100$ GeV read $(\Lambda f^2)^{1/3}>300-360$ GeV$\times\left(|c_{Xy}|\right)^{1/3}\left({\rm TeV}/{m_X}\right)^{1/6}$. XENON100 provides comparable constraints.

We stress that chiral NLO corrections might be important in certain cases~\cite{Cirigliano:2012pq}\cite{Cirigliano:2013zta}. In those instances the process $NX\to NX$ should be complemented with additional momentum-dependent terms and multi-nucleon interactions. For the cases studied here these effects are of order a few percent.

\subsection{Higgs couplings}
\label{sec:higgs}

As in the previous subsection, here all results are independent of the DM spin.

\subsubsection{Bounding Unnaturalness}

According to NDA, and assuming no specific structure in the Higgs sector, the Wilson coefficient of the operator ${\cal O}_{XH}$, coupling ${H^\dagger H}$ to DM, should be of order $\Lambda m_X/f^2$. More generally, we can write:
\ba\label{OHop}
\left.{\cal M}_T\right|_{XH}=\Delta_{\rm FT}\frac{m_h^2}{\Lambda^2}\frac{2m_X\Lambda}{f^2}\, {\bf\left(1\right)}\, \langle T'|{H^\dagger H}|T\rangle,
\ea
with $\Delta_{\rm FT}\sim\delta m_h^2/m_h^2$ the ratio between the typical contribution to the Higgs mass operator and the physical Higgs boson mass. This quantity provides a measure of the ``unnaturalness" of the theory. In the MSSM we expect $\delta m_h^2\sim\Lambda^2\sim g_w^2f^2$, with $g_w=g,g'$ a weak coupling and the new physics threshold identified with the higgsino mass. In models where the Higgs is a Nambu-Goldstone boson, however, ${\cal O}_{XH}$ is typically generated by the same (small) parameters correcting the Higgs squared mass. In this case $\delta m_h^2\sim\lambda f^2\ll\Lambda^2$, with $\lambda$ the Higgs quartic coupling.

After exchanging the Higgs boson at tree-level, from~(\ref{OHop}) we obtain the same matrix element~(\ref{Oy}) of the operator ${\cal O}_{Xy}$ -- with $y_{ij}$ identified with the quark Yukawa matrix --, but here parametrically enhanced by $\sim\Delta_{\rm FT}$. The effect of ${\cal O}_{XH}$ is therefore stronger in unnatural theories with $\Delta_{\rm FT}\gg1$, in which case ${\cal O}_{XH}$ effectively behaves like a dim-4 (dim-5) interaction for boson (fermion) DM. The current bounds on ${\cal O}_{XH}$ are readily obtained by rescaling the constraint on ${\cal O}_{Xy}$ in figure~\ref{figGauge}.

As the sensitivity of direct detection experiments improves, the relevance of~(\ref{OHop}) will get stronger {\emph{faster}} than DM signatures induced by dim-6 interactions, and eventually ${\cal O}_{XH}$ will dominate. For example, for a complex scalar, we find that the ratio between the scattering rates induced by ${\cal O}_{XH,XB}$ scales as:
\ba
\frac{\left.\sigma_T\right|_{XH}}{\left.\sigma_T\right|_{XB}}&\sim&\left(\frac{A}{Z}\right)^2\left(\frac{f_N}{\Lambda}\right)^2\left(\frac{\Lambda}{ef}\right)^4\left(\frac{\delta m_h^2}{m_h^2}\right)^2
\\\no&\sim&\left(\frac{A}{Z}\right)^2\left(\Delta_{\rm FT}\frac{v^2}{f^2}\right)^2\left(\frac{\Lambda}{20~{\rm TeV}}\right)^2,
\ea
where in the last step we used the fact that $\delta m_h^2\sim\lambda f^2$ is approximately satisfied in the scenarios of interest. This estimate tells us that (in theories compatible with NDA) the operator ${\cal O}_{XH}$ will start dominating DD signatures of complex WIMPs when experiments become sensitive to $\Lambda>O(20)$ TeV, or equivalently when they will reach sensitivities on the nucleon-DM scattering cross sections of order $10^{-47}$ cm$^2$ for $m_X\sim1$ TeV.

\subsubsection{DM and custodial $SU(2)_c$}
\label{sec:Higgs}

Complex DM can also couple via ($v_w\approx246$ GeV and $c_w$ is the cosine of the weak angle)
\ba\label{XZ}
\left.{\cal M}_T\right|_{XZ}&=&c_{XZ}\frac{2m_X}{f^2}{\bf\left(v^\mu\right)} \langle T'|H^\dagger i\overleftrightarrow{D_\mu} H|T\rangle
\\\no
&=&c_{XZ}\frac{2m_X}{f^2}{\bf \left(v^\mu\right)}\langle T'|-\frac{g}{2c_w} Z^0_\mu{v_w^2}\left(1+\frac{h}{v_w}\right)^2|T\rangle.
\ea
The analogous coupling for real DM is studied in section~\ref{sec:majo}. 

In a generic model ${\cal O}_{XZ}$ will appear in the effective action with $c_{XZ}=O(1)$. However, the coupling~(\ref{XZ}) breaks the custodial $SU(2)_c$ symmetry protecting the electro-weak $T$ parameter, so when $c_{XZ}\sim1$ important corrections to $c_w^2m_Z^2/m_W^2-1$ are expected. For example, by closing a DM loop between two insertions of ${\cal O}_{XZ}$ we obtain the effective operator $|H^\dagger D_\mu H|^2$, which immediately translates into $\alpha_{\rm em} T\sim c_{XZ}^2v_w^2/f^2$. The current $99\%$ CL bound on the electroweak $T$ parameter is on the order of $\alpha_{\rm em}|T|\lesssim2\times10^{-3}$~\cite{Baak:2012kk}, the exact value being correlated with the new physics contribution to the other oblique parameters. By imposing custodial $SU(2)_c$ the correction to $T$ can be suppressed. This may or may not be done suppressing $c_{XZ}$ as well, depending on the DM charge assignments under $SU(2)_c$ (see~\cite{Agashe:2009ja} for an explicit model). We will derive the direct detection bound for arbitrary $c_{XZ}$.

\begin{figure}
\begin{center}
\includegraphics[width=2.6in]{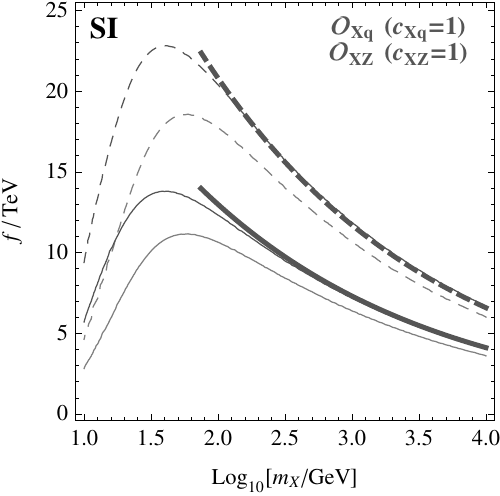}~~~~~~~~\includegraphics[width=2.75in]{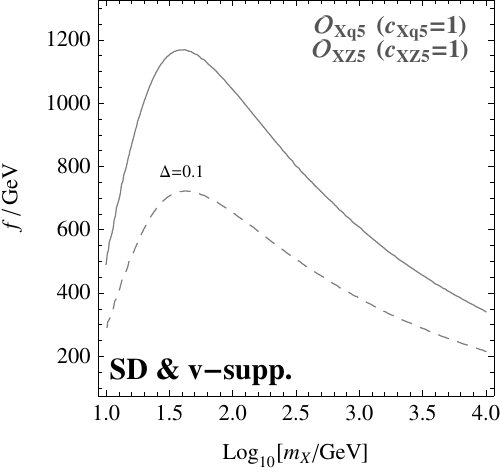}
\caption{\small Left: $90\%$ CL lower bounds on the scale $f$ suppressing the operators ${\cal O}_{XZ}$ (solid) and ${\cal O}_{Xq}$ (dashed) as a function of the DM mass for $c_{Xq,XZ}=1$. We assumed ``composite" quark doublets in ${\cal O}_{Xq}$. The upper (and darker) solid line is from LUX, the lower (lighter) from XENON100. Right: the corresponding constraints in the case of real DM (with spin).
\label{figSISD}}
\end{center}
\end{figure}

The matrix element~(\ref{XZ}) describes a coupling between DM and the $Z^0$ boson which is very severely constrained by DM experiments. Using the matching~(\ref{matchV}), and~(\ref{matchA}) including the running from $m_Z\to\mu_N$, one writes the couplings of the $Z^0$ boson to the nucleons as
\ba
\frac{g}{2c_w}Z_\mu^0\sum_{N=n,p}\overline{N}\gamma^\mu\left(\frac{c_V^{(N)}+c_A^{(N)}\gamma^5}{2}\right)N
\ea
with
\ba
c_V^{(n)}=-1,~~~~~~~~c_V^{(p)}=1-4s_w^2~~~~~~~~~c_A^{(N)}=0.85\left(-\Delta_u^{(N)}+\Delta_d^{(N)}+\Delta_s^{(N)}\right).
\ea
Finally, integrating out the $Z^0$ and performing a coherent sum of the $N$s, the spin-independent differential rate is given, neglecting small $O(v^2)$ terms, by
\ba
\left.\frac{d\sigma_T}{dE_R}\right|_{\rm SI}=\frac{\mu_{T}^2}{\pi f^4}c_{XZ}^2\left[Z(1-4s_w^2)-(A-Z)\right]^2\frac{|F(E_R)|^2}{E_R^{\rm max}}.
\ea
The spin-dependent coupling $c_A^{(N)}$ is only relevant for real DM, as discussed below. For $m_X\gtrsim100$ GeV, the LUX bound on the spin-independent rate simply reads
\ba\label{XZf}
|c_{XZ}|\frac{v_w^2}{f^2}\lesssim1.1\times10^{-3}\left(\frac{m_X}{{\rm TeV}}\right)^{1/2},
\ea
that is $f\gtrsim7.3 \sqrt{|c_{XZ}|}$ TeV$~({{\rm TeV}}/m_X)^{1/4}$. The $90\%$ CL bound from LUX (upper solid) and XENON100 (lower solid) are shown in the left of figure~\ref{figSISD} in gray. Interestingly, these are comparable to the constraint obtained from electroweak data.

\subsection{More quark and gluon operators}
\label{sec:fermions}

Stringent bounds arise if there exist direct and unsuppressed couplings between the light SM fermions and the DM sector. In these cases complex DM of arbitrary $S$ can interact with a quark current:
\ba
\left.{\cal M}_T\right|_{Xq}= c_{Xq} \frac{2m_X}{f^2}{\bf \left(v^\mu\right)}~\langle T'|{q}^\dagger\bar\sigma^\mu q|T\rangle.
\ea
The DM sector must possess a non-trivial flavor structure in order to allow this coupling and at the same time avoid problems with flavor violation. In practice, and similarly to ${\cal O}_{Xy}$, only the flavor conserving interactions are relevant to DD experiments. 

As an example, let us assume $c_{Xq}$ is a family-universal coupling for the (left-handed) quark doublet. From the matching $\sum_q{q}^\dagger\bar\sigma^\mu q\to\frac{3}{2}\overline{N}\gamma^\mu N$ (where we ignore spin-dependent interactions) we find that the LUX bound approximately reads $f\gtrsim12\times\sqrt{|c_{Xq}|}\times({\rm TeV}/m_X)^{1/4}$ TeV. The $90\%$ CL LUX and XENON100 constraints are shown in the left plot in figure~\ref{figSISD} in dashed gray. These can easily be generalized to arbitrary couplings to quark currents using~(\ref{matchV}) (\ref{matchA}).

DM could also couple to the symmetric tensors defined in eqs.~(\ref{OXqq}) (\ref{GGGG}):
\ba\label{Xqq}
\left.{\cal M}_T\right|_{Xqq, XGG}=c_{Xqq, XGG}\frac{2m_X}{\Lambda f^2}{\bf\left(v^\mu v^\nu-\frac{g^{\mu\nu}}{4}\right)}~\langle T'|O_{Xqq, XGG}^{\mu\nu}|T\rangle,
\ea
and similarly for the other tensors in~(\ref{TS}). Note that, as opposed to ${\cal O}_{Xq}$, the operator ${\cal O}_{Xqq}$ is present for both real and complex DM. Using (\ref{2S}) we get:
\ba\label{rateXqq}
\left.\frac{d\sigma_T}{dE_R}\right|_{\rm SI}=c_{Xqq, XGG}^2\frac{\mu_{T}^2}{\pi f^4}\left(\sum_{q}f^{(N)}_{q2, G2}\frac{3}{4}\frac{m_N}{\Lambda}\right)^2\frac{|F(E_R)|^2}{E_R^{\rm max}},
\ea
where we neglected terms $O(v^2)$. In the right plot of figure~\ref{figGauge} we show the bound at $90\%$ CL for $\sum_{q}f^{(N)}_{q2}=1$ (see dashed green curve). The nucleon-DM coupling is here suppressed by a factor $m_N/\Lambda$ compared to ${\cal O}_{Xq}$ because of the larger dimensionality of the operator, so the bounds are much weaker. Yet, it is {{enhanced}} by $\sim m_N/f_N$ with respect to ${\cal O}_{Xy}$, which has the same dimensionality but violates chirality. The same result holds for ${\cal O}_{XGG}$ and $\sum_{q}f^{(N)}_{G2}=1$.

Assuming the same numerical coefficients, the rate obtained from
\ba
{\bf\left(v^\mu v^\nu-\frac{g^{\mu\nu}}{4}\right)}\to{\bf\left(s^\mu s^\nu+s^\nu s^\mu+S(S+1)\frac{g^{\mu\nu}}{2}\right)}
\ea
is basically the same as in (\ref{rateXqq}) up to the replacement $\frac{3}{4}\to \frac{S(S+1)}{2}+O(v^2)$. On the other hand $v^\mu s^\nu$ gives a rate $O(v^2)$ and only provides a small correction.

\subsection{Spin-dependent Vs velocity-suppressed interactions}
\label{sec:majo}

In the case of real WIMPs, all DM vector currents in ${\cal O}_{XB,XZ,Xq}$ should be replaced by an axial-vector operator. From the explicit expression $s^\mu=(\overrightarrow{s}\cdot(\overrightarrow{v}-\overrightarrow{q}/2m_X),\overrightarrow{s})+O(v^2)$ follows that the interactions mediated by $s^\mu$ can be of two types, depending on whether the SM current dominantly matches onto $c_V\overline{N}\gamma^\mu N$ or $c_A\overline{N}\gamma^\mu\gamma^5 N$. In the non-relativistic limit the nucleon-DM hamiltonian scales as ($\overrightarrow{v_\perp}=\overrightarrow{v}-\frac{\overrightarrow{q}}{2\mu_T}$):
\ba
c_V\overrightarrow{s}\cdot\left(\overrightarrow{v_\perp}+\frac{i\overrightarrow{q}}{m_T}\wedge\overrightarrow{s_N}\right)+c_A\overrightarrow{s}\cdot\overrightarrow{s_N}+O(v^2).
\ea
The first structure leads to a velocity-suppressed, spin-independent scattering, and a rate of order $v_0^2\sim10^{-6}$ compared to the analogous vector interaction $\propto v^\mu$. The second type of interaction involves an axial nucleon current, which results in couplings proportional to the DM and the nucleon spin ($\overrightarrow{s_N}$). Currently, the most stringent bounds for spin-dependent rates are on the neutron-DM cross sections and are of the order $\sigma_n\lesssim5\times10^{-39}$ cm$^2\times m_X/{\rm TeV}$~\cite{Aprile:2013doa}, which is roughly a factor $10^{5}$ weaker than for spin-independent rates. These latter constraints can be weaker or stronger than those induced by the velocity-suppressed and  spin-independent interaction, depending on $c_A/c_V$.

Let us first discuss 
\ba\label{OXZ5}
\left.{\cal M}_T\right|_{XZ5}&=& c_{XZ5}\frac{4m_X}{f^2}{\bf \left(s^\mu\right)} \langle T'|iH^\dagger  \overleftrightarrow{D_\mu} H|T\rangle\\\no
&\to&-c_{XZ5}\frac{4m_X}{f^2}{\bf\left(s^\mu\right)} \langle T'|\overline{N}\gamma^\mu(c_V^{(N)}+c_A^{(N)}\gamma^5)N|T\rangle,
\ea
where $c^{(N)}_{V,A}$ are defined in Section~\ref{sec:Higgs}. The velocity-suppressed coupling $\propto c_V$ leads to $|c_{XZ5}|{v_w^2}/{f^2}\lesssim\sqrt{{m_X}/{{\rm TeV}}}$, that is weaker than~(\ref{XZf}) by a factor of $O(v_0)$. To estimate the effect of the spin-dependent interaction we momentarily assume $c_V=0$. Summing the spins of the final state and averaging over the initial ones, the non-relativistic neutron-DM cross section becomes
\ba
\left.\sigma_n\right|_{\rm SD}=S(S+1)\frac{4\mu_n^2}{\pi}\frac{c_{XZ5}^2}{f^4}(c^{(n)}_A)^2,
\ea
where $\mu_n$ is the DM-neutron reduced mass and $|c^{(n)}_A|\approx1.1$. The bound $\sigma_n\lesssim5\times10^{-39}$ cm$^2\times m_X/{\rm TeV}$~\cite{Aprile:2013doa} implies 
\ba
|c_{XZ5}|\frac{v_w^2}{f^2}\lesssim0.2\,\sqrt{S(S+1)}\left(\frac{m_X}{{\rm TeV}}\right)^{1/2}.
\ea
Because the axial-neutron and vector-neutron couplings are comparable, $|c^{(n)}_A|\sim |c^{(n)}_V|$, the bound from the spin-dependent interaction is found to be a factor of a few stronger than the one obtained from the velocity-suppressed interactions, as one would naively expect given the current experimental sensitivity. In the right panel of figure~\ref{figSISD} (solid line) we present the constraint with both $c^{(N)}_V, c^{(N)}_A$ turned on for $S=1/2$ DM. As anticipated, this is dominated by the spin-dependent interaction.

The conclusion is not necessarily the same for 
\ba
\left.{\cal M}_T\right|_{Xq5}= c_{Xq5}\frac{4m_X}{f^2}{\bf\left(s_\mu\right)}~\langle T'|{q}^\dagger\bar\sigma^\mu q|T\rangle,
\ea
where again $q$ refers to any quark field. The reason is that the relative magnitude of the axial and vector nucleon couplings are in this case more model-dependent. For example, the same exercise done in section~\ref{sec:fermions} for ``composite" quark doublets shows that the spin-independent constraint from LUX now gives $f\gtrsim390\times\sqrt{|c_{Xq5}|}\times({\rm TeV}/m_X)^{1/4}$ GeV (modulo a mild dependence on the DM spin). Still assuming that $q$ is the quark doublet, and that the interaction is universal for all three generations, the coupling to the axial neutron current reads ${q}^\dagger\bar\sigma^\mu q\to\Delta\overline{n}\gamma^\mu\gamma^5n$, with $\Delta\sim0.1$, see eq.~(\ref{matchA}). The resulting bound from the spin-dependent rate is weaker than the one obtained from the velocity-dependent process, since the axial coupling $c_A$ is a bit smaller than the vector coupling $c_V$. We show the combined bound (including simultaneously velocity-dependent and spin-dependent rates) on ${\cal O}_{Xq5}$ in figure~\ref{figSISD} (dashed line in the right panel) for ``composite" quark doublets, $\Delta=0.1$, and fermionic DM ($S=1/2$).

According to table~\ref{table}, the remaining DM-axial coupling would come from the anapole operator
\ba
\left.{\cal M}_T\right|_{XB5}=c_{XB5}\frac{4m_X}{\Lambda^2}{\bf\left(s^\mu\right)} \langle T'|g'\partial^\nu B_{\mu\nu}|T\rangle. 
\ea
The dominant contribution to elastic DM scattering in matter is due to photon exchange, and is spin-independent but velocity-suppressed (this expression agrees with~\cite{Gresham:2013mua}):
\ba\label{anapoleR}
\left.\frac{d\sigma_T}{dE_R}\right|_{\rm SI}=\frac{4}{3}S(S+1)\frac{\mu_{T}^2}{\pi}\left(c_{XB5}\frac{e^2Z}{\Lambda^2}\right)^2v^2\left(1-\frac{E_R}{E_R^{\rm max}}\right)\frac{|F(E_R)|^2}{E_R^{\rm max}}.
\ea
The LUX bound on $\Lambda$ is consistently suppressed by $v_0^{1/2}$ compared to that of section~\ref{sec:gauge}, apart from a mild dependence on $S$. The spin-dependent process induced by a $Z^0$-boson exchange is further suppressed by $q^2/m_Z^2\ll1$, and is therefore subleading.

\section{The un-natural pNGB Higgs: a theory of Flavor and Dark Matter}
\label{sec:theory}

Relaxing the requirement of naturalness has a crucial impact on model-building. We here discuss some of the implications in strongly coupled Higgs theories (warped extra dimensions). 

Historically, much of the efforts focused on trying to alleviate tensions with the electroweak data. However, this problem is significantly alleviated in our framework, where the compositeness scale $\Lambda$ is determined by the WIMP argument, say from a few TeV to $\Lambda\lesssim100$ TeV. In these unnatural scenarios not only is $\Lambda$ heavy enough to sufficiently suppress the S-parameter, but $f$ may be so large that the custodial symmetry $SU(2)_c$ becomes a superfluous ingredient. Similarly, one does not need a custodial parity $P_{LR}$~\cite{Agashe:2006at} to avoid large corrections to $Z^0\to b\overline{b}$. 

Having obliterated this problem, the main concern is getting the Higgs mass right. In a generic strong dynamics the Higgs has a quartic of order $\Lambda^2/f^2$, and is way too heavy. One can significantly improve by focusing on models in which the Higgs is a pseudo Nambu-Golstone boson. In this case the largest unavoidable contribution to the Higgs quartic comes from the top sector, and the Higgs potential typically scales as 
\ba
V=\frac{y_t^2N_c}{16\pi^2}\Lambda^2f^2\left[a~\sin^2\frac{h}{f}+b~\sin^4\frac{h}{f}+\dots\right],
\ea
for some coefficient $a, b$. From this expression we derive $v^2/f^2\sim a/b$ and 
\ba\label{Hmass}
m_h^2\sim(100~{\rm GeV})^2b\frac{\Lambda^2}{y_t^2f^2}.
\ea
In all known models the typical contribution to the coefficient $a$ is of order $\delta a\gtrsim1$. This has important implications: natural theories (with $\delta a\sim a$) must have $b\gtrsim1$ in order to account for $v<f$, and hence tend to predict a heavy Higgs. In order to explain the Higgs mass $m_h=125$ GeV {\emph{naturally}} one has to postulate the existence of parametrically light resonances, the so called top partners, that approximately satisfy $\Lambda/f\sim1$. In other words, natural theories for the weak scale must posses a non-trivial spectrum in order to account for the light Higgs mass.

On the other hand, $b\ll1$ is quite generic in composite Higgs models. Consider for example the symmetry breaking pattern $SU(4)\to Sp(4)$~\cite{Galloway:2010bp}\cite{Gripaios:2009pe} (delivering a Nambu-Goldstone Higgs plus a singlet), and embed the SM quarks $q$ in the $\bf{4=(2,1)\oplus(1,2)}$ of $SU(4)\supset SO(4)$ (which break $P_{LR}$). This latter scenario has the merit of being minimal, and having a very efficient collective breaking of the Higgs shift symmetry in the quark sector, as in~\cite{Vecchi:2013bja}. Moreover, the quartic $b$ arises at higher order compared to $a$: assuming that the left and right top quarks have a comparable amount of compositeness one gets 
\ba
b\sim\frac{y_t^2 f^2}{\Lambda^2}.
\ea
With this quartic, the scale and strength of the strong dynamics disappear from~(\ref{Hmass}), and a Higgs mass in the right ballpark is the {\emph{generic}} expectation. 

The lesson to be learned is that once we except some tuning of the weak scale (in this example of order $\sim\Lambda^2/m_t^2$), the Higgs mass can be obtained with a generic spectrum and even with a maximally strong sector $\Lambda\sim4\pi f$. Unnatural theories can thus appear more {\emph{generic}} in theory space.

With $\Lambda$ in the multi-TeV range the Higgs boson would be very SM-like, and colliders are not powerful, at least in the immediate future. Precision experiments appear more promising. Flavor constraints on warped 5D models, and their strongly coupled 4D dual, are known to require $\Lambda\gtrsim10$ TeV, at least in the simplest and most elegant realizations~\cite{Csaki:2008zd}\cite{Bauer:2009cf}\cite{KerenZur:2012fr}. These values are perfectly reasonable in an unnatural model. In fact, the upper bound $\sim100$ TeV suggested by the WIMP argument may be close to the reach of future high intensity experiments~\cite{Agashe:2013kxa}. Strongly coupled scenarios turn into very attractive theories for flavor~\footnote{And possibly also for the strong CP problem~\cite{Redi:2012ad}.} when pushed in the unnatural regime.

Interestingly, current DD bounds on unnatural warped extra dimensions are complementary to flavor and electroweak data. In particular, the most severe constraint from DM direct searches is on the DM electric dipole moment. Unless additional structure is introduced, this bound essentially rules out complex WIMPs with spin. The reason is that in these theories it is not generically possible to generate an $O(1)$ CKM phase and at the same time efficiently suppress CP-violation. For instance, the simplest way to relax the constraint on the DM electric dipole moment is to assume that CP is an approximate symmetry of the strong dynamics~\cite{Redi:2011zi}, and is only broken by small mixing parameters $\epsilon_q$ between the SM fermions and the strong sector. In this case the electric dipole moment arises from loop diagrams involving virtual SM fermions, and we estimate
\ba
c_{Xe}\sim\frac{\epsilon_q^2N_c\Lambda^2}{16\pi^2 f^2}.
\ea
Recalling that the top Yukawa is typically of order $y_t\sim\epsilon_q^2\Lambda/f$, we find that the relevant mass range $\Lambda\lesssim100$ TeV is still robustly excluded by LUX and XENON100, even if they constitute a fraction of the DM.

The generic expectation is that DM in these models be a scalar or a real field, with $H^\dagger H$ controlling the most promising DD signatures. Yet, in scenarios with light resonances of mass $\ll\Lambda$ some of the other operators discussed in section~\ref{sec:Xe} may be enhanced. The parametric dependence on the new physics scale $\Lambda$ of the interactions in section~\ref{sec:Xe} is expected to be accurate as long as the couplings and masses of the resonances mediating the interaction are compatible with NDA. In practice, $\Lambda$ should be interpreted as the mass of the corresponding mediator. For example, consider a vector resonance $\rho$, which can contribute to the vector-vector operators at tree-level. If it couples with NDA strength $g_\rho\sim m_\rho/f$ to the DM, while to the SM fermions only via mixing with the hypercharge gauge boson (of order $g'/g_\rho$), then the operator ${\cal O}_{XB}$ will have a coefficient as in~(\ref{XB}) except for the replacement $\Lambda^2\to m_\rho^2$.

However, a light dilaton can be an important exception to this rule, because its mass could be parametrically light in a technically natural way~\cite{Bellazzini:2013fga}\cite{Coradeschi:2013gda}, and still allow unsuppressed (NDA-sized) couplings. The dilaton $\sigma$ couples to the SM quarks, gluons, and Higgs as the Higgs boson, provided we make the replacement $h/v\to\sigma/f_\sigma$ and add model-dependent factors of order one~\cite{Goldberger:2007zk}\cite{Vecchi:2010gj}, whereas to the DM as $\sim m^2_X{X}^\dagger X\sigma/f_\sigma$. Integrating $\sigma$ out we find new contributions to the coefficients of ${\cal O}_{XG, Xy}$, which are enhanced by $\sim{\Lambda^2}/{m_\sigma^2}$ compared to the NDA expectations in~(\ref{opXG}) (\ref{Oy}) -- modulo a factor of $f^2/f_\sigma^2$. Analogously, ${\cal O}_{XH}$ may be enhanced if $m_\sigma<m_h$.

\section{Discussion}
\label{sec:con}

Naturalness is a well motivated principle, but unfortunately not a quantitative one. The new physics threshold is expected to be of order $0.1-1$ TeV, but it is not possible to predict how close to it.

An independent perspective may be provided by the WIMP ``miracle". In realistic SUSY and warped models the observed DM density is often obtained by pushing the new physics above the TeV, so the very existence of structure in the Universe may require a fine tuning of the weak scale. In this paper we showed that DM searches are an efficient probe of these unnatural theories.

We studied the direct detection of a gauge-singlet WIMP in generic extensions of the SM. We worked in a ``heavy DM" expansion, and adopted a simple unified description valid for DM of arbitrary spin $S$. DM can be real or complex, and CPT invariance restricts the couplings of the former type. The approach is fully relativistic, and explicitly manifests non-trivial correlations among the non-relativistic operators appearing in the nucleon-DM hamiltonian. For example, terms $\overrightarrow{v}^2$ are always subleading, unless cancellations occur. Similarly, $(\overrightarrow{s}\cdot\overrightarrow{v})^2$ only arises if there exist non-generic relations among irreducible Lorentz tensors. Our formalism also exploits SM gauge invariance, and should be contrasted with a $U(1)_{\rm em}\times SU(3)_{\rm color}$ invariant theory for quarks and gluons, where a larger number of relevant interactions is present.

Assuming for definiteness that the dark sector follows NDA expectations, and that CP is violated generically, the number of interactions relevant for DM direct detection reduces to those collected in table~\ref{table}. Most of the operators appear in one form or another scattered in the existing literature, which mostly focused on $S=0,1/2,1$. The only two exceptions are those associated with 
\ba
{\bf\left(s^\mu s^\nu\right)}\, \langle T'|\overline{q}\gamma^\mu iD^\nu q|T\rangle,~~~~~~~~{\bf\left(s^\mu s^\nu\right)}\, \langle T'|G^{\mu\alpha}G^{\nu}_\alpha|T\rangle,
\ea
which first arise for $S\geq1$. 

The analysis can easily be extended to more general frameworks, as well as models with inelastic DM scattering. Light mediators are accounted for by observing that these latter must be singlets under the SM group to evade collider constraints. This implies that their couplings to the SM are again given by the operators discussed in section~\ref{sec:SM}. The Wilson coefficients must now be replaced by appropriate ``form factors", which reduce to simple constants when the mediator mass is larger than the typical momentum transfer, $|\overrightarrow{q}|\lesssim100$ MeV.

The effective operators must be run from the mediator's scale down to energies at which the nucleon matrix elements are extracted. This enhances the DM couplings to the quark mass operators by $\sim20\%$, while it suppresses the axial quark currents by a similar amount.

Electromagnetic dipole moments of complex DM with spin are severely constrained by direct detection experiments, and imply important constraints on the new physics scale even when arising at loop-level. For these processes, LUX improves upon XENON100 more than for the standard spin-independent rate thanks to its better sensitivity to low recoil spectra. Interestingly, DM experiments are currently probing CP-violation at energy scales $O(10^5)$ TeV, and thus comparable to flavor experiments. We find that existing data basically rule out complex WIMPs with spin in generic strongly coupled theories for the TeV scale.

Interestingly, direct detection is complementary to electroweak precision data in the case of complex WIMPs (including scalars). The bound arising from $H^\dagger i\overleftrightarrow{D}H$ is comparable to that of the oblique $T$-parameter (and can be much stronger in certain models with custodial symmetry), whereas DM couplings to $\partial_\nu B^{\mu\nu}$ will soon be probed up to mass scales of the order of those constrained by the electroweak parameters.

For real DM of any spin the bounds on the spin-independent interactions from LUX and XENON100 are comparable, and are satisfied with a new physics threshold above the TeV. Spin-dependent interactions can compete with velocity-suppressed rates, but their relevance is in general model-dependent. For these WIMPs, processes mediated by the Higgs mass operator $H^\dagger H$ will eventually dominate over all other direct detection signatures.

\acknowledgments

We thank Kaustubh Agashe, Joachim Brod, Roberto Franceschini, Yasunori Nomura, Ian Shoemaker, Amarjit Soni, and Raman Sundrum for discussions. We acknowledge the Kavli Institute for Theoretical Physics, where this project was initiated, and the Galileo Galilei Institute, where part of this work was performed and preliminary results were presented. This work was supported in part by the NSF Grant No. PHY-0968854, No. PHY-0910467, by the Maryland Center for Fundamental Physics, and by the National Science Foundation under Grant No. PHY11-25915.

\appendix

\section{Bilinears for light DM: $S=0,1/2,1$}
\label{app:tensors}

Let us finally discuss the DM bilinears of table~\ref{tableNR}. These are derived in a $\partial/\Lambda\ll1$ expansion for $S=0,1/2,1$.~\footnote{DM bilinears for $S=3/2$ (with the exception of the 2-index symmetric tensors) are discussed in~\cite{Yu:2011by}.}. The case of scalar DM is trivial, whereas for $S=1/2,1$ the analysis is simplified by working in a small $v=|\overrightarrow{v}|$ limit. 

The wave-function $u(p)$ of a non-relativistic $S=1/2$ particle of momentum $p^\mu=mv^\mu$ reads:
\ba
u_s(p)&=&{\sqrt{m_X}}\left( \begin{array}{c}  
1-v^k\frac{\sigma^k}{2}\\
1+v^k\frac{\sigma^k}{2}
\end{array}\right)\xi_s+O(v^2),
\ea
with $\xi_s$ ($s=1,2$) a complete set of normalized 2-component spinors. The spin 3-vector is $(s^i)_{s's}=\xi_{s'}^\dagger\sigma^i\xi_{s}/2$. As explained in section~\ref{sec:bilinears}, $\overline{X}\gamma^5X$ is subleading compared to the CP-even scalar, so it is not shown in table~\ref{tableNR}. For real DM $\overline{X}\gamma^\mu X=0$, and $\overline{X}\partial^\mu X$ is a total derivative, while $\overline{X}\sigma^{\mu\nu}i\partial_\nu X$ and $\overline{X}\sigma^{\mu\nu}\gamma^5\partial_\nu X$ are subleading compared to $\partial^\mu(\overline{X} X), \overline{X}\gamma^\mu\gamma^5X$.

According to the general argument discussed in the text, the anti-symmetric tensors for Majorana $X$ are $\propto q$. For example, from the identify $\overline{X}\gamma^{\mu}\gamma^5\partial^{\nu} X=\frac{1}{2}\partial^\nu(\overline{X}\gamma^{\mu}\gamma^5X)$ follows that $\overline{X}\gamma^{[\mu}\gamma^5\partial^{\nu]} X$ can be integrated by parts to give a coupling for $\overline{X}\gamma^{\mu}\gamma^5X$. Similarly, the differential rate of $\overline{X}\gamma^{\mu} i\partial^{\nu}Xg'\widetilde B_{\mu\nu}/\Lambda^2$ is the same as that obtained from the anapole operator ${\cal O}_{XB5}$. For Dirac DM $\overline{X}(\gamma^{\mu} i\partial^{\nu}-\gamma^{\nu} i\partial^{\mu}) X$ gives a correction $O(q/\Lambda)\ll1$ to the electric dipole and $O(m_X/\Lambda)$ to the magnetic dipole.

Next, we describe the spin-1 DM by a Proca field with $\partial_\mu X^\mu=0$. The polarization vectors in the rest frame can be chosen to be $\epsilon^\mu_s=(0,\delta_s^i)$. Making a Galilean boost, we find ($p^\mu=m_Xv^\mu$)
\ba
\epsilon^\mu_s(p)=\left( \begin{array}{c}  
v^k\epsilon^k_s\\
\epsilon^i_s
\end{array}\right)+O(v^2),
\ea
with $\epsilon^\mu_s(p)\epsilon_{\mu s'}(p)=\epsilon^\mu_s\epsilon_{\mu s'}=-\delta_{ss'}$ and $\sum_s\epsilon_s^\mu(p)\epsilon^\nu_s(p)=-(g^{\mu\nu}-v^\mu v^\nu)$. The $3$-dimensional representation of the spin operator is $(s^i)_{s's}=i\varepsilon^{ijk}\epsilon_{s'}^j\epsilon^k_s$, with $i\varepsilon^{ijk}$ the infinitesimal generators of rotations. The 4-spin operator may be written as $s^\mu=-i\varepsilon^{\mu\nu\alpha\beta}v_\nu\epsilon_{\alpha,s'}\epsilon_{\beta,s}$.

For the same reasons explained for the spin-0 case we ignored $\partial^\mu(\overline{X_\alpha}X^\alpha)$. Similarly, we did not not show $\overline{X_\alpha}\partial^\alpha X^\mu=\partial^\alpha(\overline{X_\alpha}X^\mu)$.

In the 2-index bilinears there often appears the symmetric tensor $s^{ij}=s^is^j+s^js^i-2\delta^{ij}$ (note that $s^{kk}=2+2-2\times3\neq0$). The following identity is useful:
\ba\label{sij}
\epsilon_{s'}^i\epsilon_s^j=-\frac{i}{2}\varepsilon^{ijk}(s^k)_{s's}-\frac{1}{2}(s^{ij})_{s's}.
\ea
Making use of~(\ref{sij}) one may find an expression for the 2-index tensors. The $O(q)$ terms in these tensors are in general different from those of $S=1/2$ DM. All anti-symmetric tensors at the 2-derivative level vanish for $q=0$.

\section{$\langle N|GG|N\rangle$ and $\langle N|m_q\overline{q}q|N\rangle$ at NLO}
\label{sec:RG}

Consider the following Lagrangian
\ba
{\cal L}^{Gq}_{\rm eff}= C_G G_{\mu\nu}^aG_{\mu\nu}^a+\sum_{q=1}^{N_f} C_q m_q\overline{q}q.
\ea
The operators in ${\cal L}^{Gq}_{\rm eff}$ are renormalized at some RG scale $\Lambda$, with the $C_i$s obtained from matching with the UV theory at this scale.~\footnote{This was referred to as a ``short-distance effect" in~\cite{Hisano:2010ct}.} 

We work at leading order in $C_i$ and the weak couplings, so all RG effects are induced by the strong coupling $\alpha_s=g_s^2/4\pi$. In a mass-independent renormalization scheme, the quark mass operator is multiplicatively renormalized, $\frac{d}{d\mu}(m_q\overline{q}q)=0$, whereas $GG$ mixes with $m_q\overline{q}q$. One way to obtain its RG evolution is to recall that the matrix elements of the QCD scale anomaly~\cite{Collins:1976yq}
\ba\label{anomaly}
\langle {\rm out}|\left.\theta^\mu_\mu\right|_{\rm QCD}|{\rm in}\rangle=\langle {\rm out}|\gamma GG+(1-\gamma_m)\sum_{q=1}^{N_f}m_q\overline{q}q|{\rm in}\rangle,~~~~~~~~\left(\gamma\equiv \frac{\beta_s}{2g_s}\right)
\ea
are RG invariant. Here $\beta_s=\partial g_s/\partial\log\mu$ and $\gamma_m=+\partial\log m_q/\partial\log\mu$ (our convention is such that the dimension of the mass operator $\overline{q}q$ is $3+\gamma_m$ and ${\cal L}_{\rm QCD}\supset-\sum_qm_q\overline{q}q$ is the quark mass term). Requiring $\frac{d}{d\mu}\langle {\rm out}|\left.\theta^\mu_\mu\right|_{\rm QCD}|{\rm in}\rangle=0$ gives:~\footnote{Strictly speaking, the operators $m_q\overline{q}q,GG$ do not form a closed set under renormalization. However, the other operators ($E_i$) they mix with vanish on-shell, so they do not contribute to the matrix element ($\langle N'|E_i|N\rangle=0$), and our argument is unchanged. In other words, the RG evolution of the $C_i$s decouples from those of the operators vanishing on-shell (see~\cite{Jenkins:2013zja} for a recent discussion).} 
\ba\label{RGGG}
\mu\frac{d}{d\mu}\left( \begin{array}{c}  
GG\\
m_q\overline{q}q
\end{array}\right)=\frac{\beta_s}{\gamma}\left( \begin{array}{cc}  
-\gamma' & \gamma'_m\\
0 & 0
\end{array}\right)\left( \begin{array}{c}  
GG\\
m_q\overline{q}q
\end{array}\right),
\ea
(primes indicate derivatives with respect to the renormalized coupling $g_s$) and agrees with the RG evolution found in~\cite{Grinstein:1988wz}, where no explicit use of the non-renormalization of the scale anomaly was made. Independence of the effective Lagrangian on the renormalization point provides an RG equation for the coefficients $C_i$. The exact solution is
\ba\label{RG}
C_{G}(\mu)&=&\frac{\gamma(\mu)}{\gamma(\mu_0)}C_{G} (\mu_0)\\\no
C_{q}(\mu)&=&C_{q}(\mu_0)-\left(\frac{\gamma_m(\mu)-\gamma_m(\mu_0)}{\gamma(\mu_0)}\right)C_{G}(\mu_0).
\ea

At a heavy quark threshold $\mu=m_{q_h}$ the theory should be matched onto an effective Lagrangian for the light degrees of freedom. This latter schematically reads $C_G G_{\mu\nu}^aG_{\mu\nu}^a+\sum_{q\neq q_{h}} C_q m_q\overline{q}q+O(1/m_{q_h}^2)$, where the last term refers to higher dimensional operators suppressed by $m_{q_h}$. At 1-loop order, only two dimension-6 operators are generated~\cite{Cho:1994yu}
\ba\label{G6}
-\frac{\alpha_s(m_{q_h})}{60\pi m_{q_h}^2}(DG)^a_{\mu}(DG)^a_{\mu}-\frac{\alpha_s(m_{q_h})}{720\pi m_{q_h}^2}g_sf_{abc}G^a_{\mu\nu}G^b_{\mu\rho}G^c_{\nu\rho}.
\ea
As a measure of their impact we can estimate the correction induced by~(\ref{G6}) on the nucleon matrix element. Parametrizing the RG effects $m_{q_h}\to \mu_N\sim m_N$ by a renormalization of the coupling $g_s$, and using NDA, we expect $O\left({1}/{m_{q_h}^2}\right)\sim[{\alpha_s(\mu_N)}/{60\pi}]{m_N^2}/{m_{q_h}^2}$. By far the largest correction comes from the charm quark. Yet, even taking $\alpha_s(\mu_N)\sim1$ the latter correction amounts to just a few percent. In the following we will therefore work at leading order in a heavy quark expansion. 

Now, in order for both ${\cal L}^{Gq}_{\rm eff}$ and~(\ref{anomaly}) to be RG invariant there must appear a discontinuity at $\mu=m_{q_h}$:
\ba
(1-\gamma_m^+)m_{q_h}\overline{q_h}q_h&=&[\gamma^--\gamma^+]GG\\\no
&+&\sum_{q=1}^{N_f-1}[\gamma_m^+-\gamma_m^-]m_q\overline{q}q+O\left(\frac{1}{m_{q_h}^2}\right).
\ea
Here $^{+(-)}$ indicates that the function is evaluated at $m^{+(-)}_{q_h}$, with the appropriate number of active flavors. In the low energy theory the above discontinuity is parametrized by a matching condition on the Wilson coefficients:
\ba\label{match}
C_G(m^-_{q_h})&=&C_G(m^+_{q_h})+\frac{\gamma^--\gamma^+}{1-\gamma_m^+}C_{q_h}(m^+_{q_h})+O\left(\frac{1}{m_{q_h}^2}\right)\\\no&=&C_G(m^+_{q_h})-\frac{\alpha_s(m_{q_h})}{12\pi}\left(1+{11}\frac{\alpha_s(m_{q_h})}{4\pi}+O(\alpha_s^2)\right)C_{q_h}(m_{q_h}^+)+O\left(\frac{1}{m_{q_h}^2}\right)\\\no
C_{q}(m^-_{q_h})&=&C_{q}(m^+_{q_h})+\frac{\gamma_m^+-\gamma_m^-}{1-\gamma_m^+}C_{q_h}(m^+_{q_h})+O\left(\frac{1}{m_{q_h}^2}\right)\\\no&=&C_{q}(m^+_{q_h})+O\left(\alpha_s^2, \frac{1}{m_{q_h}^2}\right).
\ea
Observe that eq.~(\ref{match}) reduces at 1-loop to the formula of~\cite{Shifman:1978zn}. The heavy quark renormalizes the mass operators of the light quarks starting from two-loop diagrams. We estimate that $({\gamma_m^+-\gamma_m^-})/({1-\gamma_m^+})\leq O(1-2\%)$, so this latter effect is always rather small.

Ultimately, we want an effective field theory for the nucleons. Using our definition (\ref{fTqdef}), and recalling that $\langle N|\theta_{\mu\nu}|N\rangle=2k_\mu k_\nu$ (with $N=n,p$ a nucleon), we get: 
\ba\label{FN}
\frac{\langle N|{\cal L}^{Gq}_{\rm eff}|N\rangle}{2m^2_N}=\sum_{q=u,d,s,c,b,t}f_{Tq}^{(N)}C_q(\Lambda)+\frac{7}{9}\frac{C_G(\Lambda)}{\gamma(\Lambda)}f_{TG}^{(N)}(\Lambda).
\ea

One can derive analytical expressions for $f_{Tq,TG}^{(N)}$ by turning on one $C_i(\Lambda)$ at a time and employing~(\ref{RG}) and~(\ref{match}):
\ba\label{fTc}
f_{Tc}^{(N)}&=&
\frac{\gamma(m_c^-)-\gamma(m_c^+)}{\gamma(m^-_c)[1-\gamma_m(m_c^+)]}\left[1-(1-\gamma_m(m^-_{c}))\sum_{q=u,d,s}f_{Tq}^{(N)}\right]\\\no
&+&\frac{\gamma_m(m_c^+)-\gamma_m(m_c^-)}{1-\gamma_m(m_c^+)}\sum_{q=u,d,s}f_{Tq}^{(N)}+O\left(\frac{1}{m_c^2}\right)\\\no
&=&\frac{2}{27}\left[1+\frac{35}{9}\frac{\alpha_s(m_c)}{4\pi}-\left(1+\frac{107}{9}\frac{\alpha_s(m_c)}{4\pi}\right)\sum_{q=u,d,s} f_{Tq}^{(N)}\right]+O\left(\alpha_s^2,\frac{1}{m_{c}^2}\right),
\ea
and similar expressions (though more involved) for the other $f_{Tq}^{(N)}$s. For example, neglecting $f_{Tu,Td,Ts}^{(N)}$:
\ba\label{fTapp}
f_{Tb}^{(N)}\sim\frac{\gamma(m_b^-)-\gamma(m_b^+)}{1-\gamma_m(m_b^+)}\times\frac{\gamma(m_c^+)}{\gamma(m^-_b)}\times\frac{1}{\gamma(m_c^-)},
\ea
and analogously for $f_{Tt}^{(N)}$. The first factor in~(\ref{fTapp}) originates from matching $m_b\overline{b}b$ onto $GG(m^-_b)$, the second from the running $m_b^-\to m_c^+$, and the third from the matching $GG\to\Theta^\mu_\mu/\gamma$ at $m_c$ (valid only in the present limit $f_{Tu,Td,Ts}^{(N)}=0$).

To proceed with our numerical analysis we use the 4-loop QCD beta-function and mass anomalous dimension in the ${\overline{\rm MS}}$ scheme~\cite{vanRitbergen:1997va}, expand the analytic expressions of $f^{(N)}_{Tq,TG}$ up to $O(\alpha_s^3)$, and assume $\alpha_s(m_Z)=0.1184$. The values of the running quark masses $m_q(m_q)$, with errors, are taken from the Particle Data Group~\cite{PDG}. The uncertainties in the quark masses result in:~\footnote{We stress that $\alpha_s$ in a generic mass-independent scheme is discontinuous at threshold. The discontinuity is accounted for by introducing appropriate matching functions, collected for instance in~\cite{Chetyrkin:2000yt}. However, this effect is well within the mass errors, so we can safely take $\alpha_s(m_{q_h}^+)=\alpha_s(m_{q_h}^-)$ in what follows.}
\ba
\alpha_s(m_t)&=&0.108,\\\no
\alpha_s(m_b)&=&0.226- 0.227,\\\no
\alpha_s(m_c)&=&0.387- 0.398.
\ea
We find the results shown in (\ref{result2'}) as well as (see also~(\ref{result3'})):
\ba\label{result3}
f_{TG}^{(N)}(\Lambda)
&=&\left(0.97-0.93\sum_{q=u,d,s}f_{Tq}^{(N)}\right)+O\left(\frac{g^2_s(\Lambda)}{16\pi^2}\frac{1}{m_{c}^2}\right).
\ea
The one-loop result is recovered by setting to unity all numerical coefficients in the parenthesis of~(\ref{result2'}) (\ref{result3}). The uncertainty in the quark masses has basically no effect on the final result, showing a very mild dependence on the RG scale. The convergence in the perturbative expansion is rather satisfactory, despite the low RG scale, with N$^{n+1}$LO/N$^{n}$LO$=O(10\%,40\%,20\%)$.

The size of the NLO corrections in (\ref{result2'}) is of order $\alpha_sN_c/4\pi$, so the larger effect is consistently found in $f_{Tc}^{(N)}$ and is $O(10-20\%)$.~\footnote{The factor of $(\gamma(m_c^-)-\gamma(m_c^+))/(1-\gamma_m(m_c^+))$ in eq.~(\ref{match}) was approximated at the 2-loop level in ref.~\cite{Djouadi:2000ck}, but $1/\gamma(m_c^-)$ was erroneously taken to be at one-loop order. This overestimates the induced $C_G$, and hence $f_{Tq}^{(N)}$.} The lattice simulations of~\cite{Freeman:2012ry}\cite{Gong:2013vja} are compatible with our $f_{Tc}^{(N)}$. The NLO corrections in (\ref{result3}) are also small, and are dominated by~(\ref{RG}).

We conclude by noting that we find good agreement with~\cite{Kryjevski:2003mh} on the size of the NLO corrections in~(\ref{result2'}) (though only in the limit $m_u=m_d=m_s=0$). However, the leading expressions shown in eq.~(1) of~\cite{Kryjevski:2003mh} do not reproduce the $2/27$ in eq.~(\ref{result2'}). What is missing are terms analogous to the last factor in~(\ref{fTapp}).


 \end{document}